\documentclass [preprint,showpacs,nofootinbib,superscriptaddress]{revtex4}
\usepackage{bbm}
\usepackage{verbatim}
\usepackage{slashed}
\usepackage{epsfig}
\usepackage{graphicx}
\usepackage{amsmath}
\usepackage{mathrsfs}
\usepackage{bm}

\newcommand{\re}{\mathop{\mathrm{Re}}\nolimits}

\begin{document}

\preprint{DESY~14-246\hspace{12cm}ISSN~0418-9833}
\boldmath
\title{Relativistic corrections to prompt $J/\psi$ photo- and hadroproduction}
\unboldmath


\author{Zhi-Guo He and Bernd A.~Kniehl}

\affiliation{
{\normalsize II. Institut f\"ur Theoretische Physik, Universit\"at Hamburg,}\\
{\normalsize Luruper Chaussee 149, 22761 Hamburg, Germany}
}


\date{\today}

\begin{abstract}\vspace{5mm}
We systematically calculate the relativistic corrections to prompt $J/\psi$
photoproduction and hadroproduction using the factorization formalism of
nonrelativistic QCD.
Specifically, we include the ${}^3\!S_1^{[1]}$ and ${}^3\!P_{J}^{[1]}$
color-singlet and the ${}^3\!S_1^{[8]}$, ${}^1\!S_{0}^{[8]}$, and
${}^3\!P_{J}^{[8]}$ color-octet channels as well as the effects due to the
mixing between the ${}^3\!S_1^{[8]}$ and ${}^3\!D_1^{[8]}$ channels.
We provide all the squared hard-scattering amplitudes in analytic form.
Assuming the nonrelativistic-QCD long-distance matrix elements to satisfy the
velocity scaling rules, we find the relativistic corrections to be appreciable,
except in the ${}^3\!S_1^{[1]}$ color-singlet channel of hadroproduction.
We also observe significant differences in the line shapes of the relativistic
corrections between photoproduction and hadroproduction.

\end{abstract}


\maketitle
\section{Introduction}

The production of heavy quarkonia, the QCD bound states of heavy-quark pairs
($Q\bar{Q}$), serves a an ideal laboratory to probe both the perturbative and
nonperturbative aspects of QCD.
The effective quantum field theory of nonrelativistic QCD (NRQCD)
\cite{Caswell:1985ui} endowed with the factorization formalism introduced by
Bodwin, Braaten, and Lepage \cite{Bodwin:1994jh} is nowadays the most
favorable theoretical approach to study heavy-quarkonium production and decay.
In this framework, the theoretical predictions are separated into
process-dependent short-distance coefficients (SDCs) and supposedly universal
long-distance matrix elements (LDMEs).
The SDCs may be calculated perturbatively as expansions in the strong-coupling
constant $\alpha_{s}$, while the LDMEs are predicted to scale with definite
powers of the relative velocity $v$ of the heavy quarks in the quarkonium rest
frame \cite{Lepage:1992tx}.
In this way, the theoretical calculations are organized as double expansions in
$\alpha_s$ and $v$.
In contrast to the color-singlet (CS) model, in which the $Q\bar{Q}$ pair must
be in the CS state that shares the spin $S$, orbital angular momentum $L$, and
total angular momentum $J$ with the considered quarkonium, NRQCD accommodates
all possible Fock states $n={}^{2S+1}\!L_{J}^{[a]}$, where $a=1,8$ stands for
CS and color octet (CO), respectively.

During the past two decades, tests of NRQCD factorization and the universality
of the LDMEs were performed in a vast number of experimental and theoretical
works.
Despite numerous great successes, there are still some challenges in $J/\psi$
hadroproduction.
As for the prompt yield of unpolarized $J/\psi$ mesons, the
next-to-leading-order (NLO) QCD corrections were calculated for all the
relevant channels, including ${}^3\!S_1^{[1]}$ \cite{Campbell:2007ws},
${}^3\!S_1^{[8]}$, ${}^1\!S_0^{[8]}$ \cite{Gong:2008sn}, and ${}^3\!P_{J}^{[8]}$
\cite{Butenschoen:2010rq,Ma:2010yw} for direct production and the feed down
from $\psi^\prime$ mesons, and
${}^3\!P_{J}^{[1]}$ \cite{Ma:2010vd,Butenschoen:2013pxa}
for the feed down from $\chi_{cJ}$ mesons.
As for the LDMEs, different ways of fitting lead to different results.
By a global fit to the world's $J/\psi$ data from hadroproduction,
photoproduction, two-photon scattering, and $e^{+}e^{-}$ annihilation, the
three CO LDMEs relevant for direct production were successfully pinned down
\cite{Butenschoen:2011yh} in a way compatible with the velocity scaling rules
\cite{Lepage:1992tx}, which greatly supported their universality.
Using these LDMEs, the $J/\psi$ polarization in hadroproduction was predicted
to be largely transversal \cite{Butenschoen:2012px}.
By fitting to the CDF Run II measurements of $J/\psi$ yield and polarization
for transverse momenta $p_T>7~\mathrm{GeV}$, the authors of
Ref.~\cite{Chao:2012iv} obtained two linear combinations of the three LDMEs,
which led to an almost unpolarized prediction for the LHC.
Fitting to the prompt $J/\psi$ yield measured for $p_T>7~\mathrm{GeV}$ by
CDF~II and LHCb, including also the feed-down contributions from the
$\chi_{cJ}$ and $\psi^\prime$ mesons, a third set of LDMEs was obtained, which
resulted in a moderately transverse  $J/\psi$ polarization at the LHC
\cite{Gong:2012ug}.
All of these three LDME sets can describe well the $J/\psi$ yield at the LHC.
Unfortunately, none of the resulting predictions for $J/\psi$ polarization is
consistent with the latest measurements by CMS \cite{Chatrchyan:2013cla} and
LHCb \cite{Aaij:2013nlm}.
The above results reflect the fact that the NRQCD prediction of prompt $J/\psi$
polarization strongly depends on the actual values of the LDMEs.
To test the universality of the LDMEs in a meaningful way and to clarify the
$J/\psi$ polarization puzzle, it is useful to investigate some other effects in
the determination of the LDMEs, such as the higher-order relativistic
corrections.

In the heavy-quarkonium system, we actually have $v^2\sim \alpha_{s}(2m_{Q})$,
which is not very small.
In some cases, it was found that the higher-order $v^2$ corrections are even as
important as the higher-order $\alpha_s$ corrections.
For example, the relativistic corrections played an important role in resolving
both the double-charmonium \cite{He:2007te} and $J/\psi+X_{\mathrm{non-cc}}$
production problems at the $B$ factories \cite{He:2009uf,Jia:2009np}.
In $J/\psi$ hadroproduction, the $v^2$ corrections in the CO channels
${}^1\!S_0^{[8]}$ and ${}^3\!S_1^{[8]}$ were found to be significant in the
large-$p_T$ region \cite{Xu:2012am}, although they are tiny in the CS
${}^3\!S_1^{[1]}$ channel \cite{Fan:2009zq}.
In double-quarkonium hadroproduction, the $v^2$ corrections also turned out to
be significant \cite{Martynenko:2012tf}, especially in the CO channels
\cite{Li:2013csa}.
In the test of the hypothesis $X(3872)=\chi_{c1}^\prime$ in hadroproduction,
appreciable $v^2$ corrections were encountered in the ${}^3\!P_1^{[1]}$ channel
\cite{Butenschoen:2013pxa}.
All these observations provide a strong motivation for us to systematically
study the $v^2$ corrections to the cross sections of prompt $J/\psi$
photoproduction and hadroproduction.
This will allow us to render global fits of the contributing LDMEs more
reliable and to deepen our understanding of their universality.
While the SDCs of direct $J/\psi$ production immediately carry over to the
feed down from the $\psi^\prime$ mesons, the feed down from the $\chi_{cJ}$
mesons requires a separate calculation.

The remainder of this paper is organized as follows.
In Sec.~\ref{sec:two}, we explain how we calculate all the relevant SDCs.
Our numerical results are presented in Sec.~\ref{sec:three}.
Our conclusions are contained in Sec.~\ref{sec:four}.
Our analytic results are listed in the Appendix.

\section{NRQCD factorization formula}
\label{sec:two}

Invoking the Weizs\"{a}cker--Williams approximation and the factorization
theorem of the QCD parton model, the cross sections for the photoproduction or
hadroproduction of the hadron $H=J/\psi,\chi_{cJ},\psi^\prime$ may be written
as
\cite{Butenschoen:2011yh}
\begin{equation}\label{xs}
\sigma(AB\to H+X)
=\sum_{i,j,k} \int dx_1 dy_1 d x_2\, f_{i/A}(x_1)f_{j/i}(y_1)f_{k/B}(x_2)
\hat{\sigma}(jk\to H +X),
\end{equation}
where $f_{i/A}(x)$ is the parton distribution function (PDF) of the parton $i$
in the hadron $A=p,\bar{p}$ or the flux function of the photon $i=\gamma$ in
the charged lepton $A=e^-,e^+$, $f_{j/i}(y_1)$ is $\delta_{ij}\delta(1-y_1)$ or
the PDF of the parton $j$ in the resolved photon $i=\gamma$, and
$\hat{\sigma}(jk\to H+X)$ is the partonic cross section.
In NRQCD through relative order $v^2$, the latter is factorized as
\cite{Bodwin:1994jh}
\begin{equation}
\hat{\sigma}(ij\to H +X)=
\sum_{n}\left(
\frac{F_{ij}(n)}{m_c^{d_{\mathcal{O}(n)}-4}}\langle\mathcal{O}^{H}(n)\rangle+
\frac{G_{ij}(n)}{m_c^{d_{\mathcal{P}(n)}-4}}\langle\mathcal{P}^{H}(n)\rangle
\right),
\label{eq:fg}
\end{equation}
where $\mathcal{O}^H(n)$ is the four-quark operator pertaining to the
transition $n\to H$ at leading order (LO) in $v$, with mass dimension
$d_{\mathcal{O}(n)}$; $\mathcal{P}^H(n)$ is related to its $v^2$ correction and
carries mass dimension $d_{\mathcal{P}(n)}=d_{\mathcal{O}(n)}+2$; and $F_{ij}(n)$ and
$G_{ij}(n)$ are the appropriate SDCs of the partonic subprocesses
$i+j\to c\bar{c}(n)+X$.
Working in the fixed-flavor-number scheme, the parton $i$ runs over the gluon
$g$ and the light quarks $q=u,d,s$ and antiquarks $\bar{q}$.

According to the velocity scaling rules \cite{Lepage:1992tx}, the leading
contributions to direct $J/\psi$ and $\psi^\prime$ production are due to the
${}^3\!S_1^{[1]}$, ${}^3\!S_1^{[8]}$, ${}^1\!S_0^{[8]}$, and ${}^3\!P_J^{[8]}$
channels, and those to direct $\chi_{cJ}$ production are due to the
${}^3\!P_J^{[1]}$ and ${}^3\!S_1^{[8]}$ channels.
Accordingly, prompt $J/\psi$ photoproduction and hadroproduction proceeds at LO
through the partonic subprocesses
\begin{eqnarray}
g+\gamma &\to& c\bar{c}({}^3\!S_1^{[1,8]},{}^1\!S_0^{[8]},{}^3\!P_J^{[8]})+g,
\nonumber\\
q(\bar{q})+\gamma &\to& c\bar{c}({}^3\!S_1^{[8]},{}^1\!S_0^{[8]},{}^3\!P_J^{[8]})+q(\bar{q}),
\nonumber\\
g+g &\to& c\bar{c}({}^3\!S_1^{[1,8]},{}^1\!S_0^{[8]},{}^3\!P_J^{[1,8]})+g,
\nonumber\\
q(\bar{q})+g &\to& c\bar{c}({}^3\!S_1^{[8]},{}^1\!S_0^{[8]},{}^3\!P_J^{[1,8]})+q(\bar{q}),
\nonumber\\
\bar{q}+q &\to& c\bar{c}({}^3\!S_1^{[8]},{}^1\!S_0^{[8]},{}^3\!P_J^{[1,8]})+g.
\label{eq:sub}
\end{eqnarray}
We adopt the definitions of the relevant four-quark operators
$\mathcal{O}^H(n)$ from the literature \cite{Bodwin:1994jh,Brambilla:2008zg}
and define the corresponding four-quark operators $\mathcal{P}^H(n)$ as:
\begin{eqnarray}
\mathcal{P}^{J/\psi}(^{3}S_{1}^{[1]})&=&\chi^{\dagger}\boldsymbol{\sigma}^{i}\psi
(a^{\dagger}_{J/\psi}a_{J/\psi})\psi^{\dagger}\boldsymbol{\sigma}^{i}\left(-\frac{i}{2}
\overleftrightarrow{\boldsymbol{D}}\right)^{2}\chi+{\rm H.c.},
\nonumber\\
\mathcal{P}^{J/\psi}(^{1}S_{0}^{[8]})&=&\chi^{\dagger}T^{a}\psi
(a^{\dagger}_{J/\psi}a_{J/\psi})\psi^{\dagger}T^{a}\left(-\frac{i}{2}
\overleftrightarrow{\boldsymbol{D}}\right)^{2}\chi+{\rm H.c.},
\nonumber\\
\mathcal{P}^{J/\psi}(^{3}P_{J}^{[8]})&=&\chi^{\dagger}\boldsymbol{\sigma}^{i}
\left(-\frac{i}{2}\overleftrightarrow{\boldsymbol{D}^{j}}\right)T^{a}\psi
(a^{\dagger}_{J/\psi}a_{J/\psi})\psi^{\dagger}\boldsymbol{\sigma}^{i}T^{a}
\left(-\frac{i}{2}\overleftrightarrow{\boldsymbol{D}^{j}}\right)
\left(-\frac{i}{2}
\overleftrightarrow{\boldsymbol{D}}\right)^{2}\chi+{\rm H.c.},
\nonumber\\
\mathcal{P}^{\chi_{c0}}(^{3}P_{0}^{[1]})&=&\frac{1}{3}\chi^{\dagger}
\left(-\frac{i}{2}\overleftrightarrow{\boldsymbol{D}}\cdot\boldsymbol{\sigma}
\right)\psi
(a^{\dagger}_{\chi_{c0}}a_{\chi_{c0}})\psi^{\dagger}\left(-\frac{i}{2}
\overleftrightarrow{\boldsymbol{D}}\cdot\boldsymbol{\sigma}\right)
\left(-\frac{i}{2}
\overleftrightarrow{\boldsymbol{D}}\right)^{2}\chi+{\rm H.c.},
\nonumber\\
\mathcal{P}^{\chi_{c1}}(^{3}P_{1}^{[1]})&=&\frac{1}{2}\chi^{\dagger}
\left(-\frac{i}{2}\overleftrightarrow{\boldsymbol{D}}\times\boldsymbol{\sigma}
\right)\psi
(a^{\dagger}_{\chi_{c1}}a_{\chi_{c1}})\psi^{\dagger}\left(-\frac{i}{2}
\overleftrightarrow{\boldsymbol{D}}\times\boldsymbol{\sigma}\right)
\left(-\frac{i}{2}
\overleftrightarrow{\boldsymbol{D}}\right)^{2}\chi+{\rm H.c.},
\nonumber\\
\mathcal{P}^{\chi_{c2}}(^{3}P_{2}^{[1]})&=&\frac{1}{2}\chi^{\dagger}
\left(-\frac{i}{2}\overleftrightarrow{\boldsymbol{D}}^{(i}\boldsymbol{\sigma}^{j)}
\right)
\psi(a^{\dagger}_{\chi_{c2}}a_{\chi_{c2}})\psi^{\dagger}\left(-\frac{i}{2}
\overleftrightarrow{\boldsymbol{D}}^{(i}\boldsymbol{\sigma}^{j)}\right)
\left(-\frac{i}{2}
\overleftrightarrow{\boldsymbol{D}}\right)^{2}\chi+{\rm H.c.},
\nonumber\\
\mathcal{P}^{H}(^{3}S_{1}^{[8]})&=&\chi^{\dagger}\boldsymbol{\sigma}^{i}T^{a}\psi
(a^{\dagger}_{H}a_{H})\psi^{\dagger}\boldsymbol{\sigma}^{i}T^{a}\left(-\frac{i}{2}
\overleftrightarrow{\boldsymbol{D}}\right)^{2}\chi+{\rm H.c.},
\nonumber\\
\mathcal{P}^{H}(^{3}S_{1}^{[8]},^{3}D_{1}^{[8]})&=&\sqrt{\frac{3}{5}}\chi^{\dagger}
\sigma^{i}T^{a}\psi (a^{\dagger}_{H}a_{H})\psi^{\dagger}\boldsymbol{\sigma}^{j}
\boldsymbol{K}^{ij}T^{a}\chi+{\rm H.c.},
\label{eq:pop}
\end{eqnarray}
where
$\overleftrightarrow{\boldsymbol{D}}^{(i}\boldsymbol{\sigma}^{j)}
=(\overleftrightarrow{\boldsymbol{D}}^{i}\boldsymbol{\sigma}^{j}
+\overleftrightarrow{\boldsymbol{D}}^{j}\boldsymbol{\sigma}^{i})/2
-\overleftrightarrow{\boldsymbol{D}}\cdot\boldsymbol{\sigma}\delta^{ij}/3$ and
$\boldsymbol{K}^{ij}
=(-i/2)^2(\overleftrightarrow{\boldsymbol{D}}^{i}
\overleftrightarrow{\boldsymbol{D}}^{j}
-\overleftrightarrow{\boldsymbol{D}}^{2}\delta^{ij}/3)$.
Our definition of the $S$-$D$ mixing operator
$\mathcal{P}^{H} (^{3}S_{1}^{[8]},^{3}D_{1}^{[8]})$ differs from that in
Refs.~\cite{Bodwin:1994jh,Brambilla:2008zg}, where a linear combination of
$\mathcal{P}^{H}(^{3}S_{1}^{[8]})$ and
$\mathcal{P}^{H}(^{3}S_{1}^{[8]},^{3}D_{1}^{[8]})$ in Eq.~(\ref{eq:pop}) is
used instead.
At order $v^2$, there are no heavy-quark spin symmetries among the
$\mathcal{O}^H(n)$ operators, but they still hold among the $\mathcal{P}^H(n)$
operators, yielding the relationships
\begin{eqnarray}
\langle\mathcal{P}^{J/\psi}(^{3}P_{0}^{[8]})\rangle
&=&\frac{1}{2J+1}\langle\mathcal{P}^{J/\psi}(^{3}P_{J}^{[8]})\rangle,
\nonumber\\
\langle\mathcal{P}^{\chi_{c0}}(^{3}S_{1}^{[8]})\rangle
&=&\frac{1}{2J+1}\langle\mathcal{P}^{\chi_{cJ}}(^{3}S_{1}^{[8]})\rangle,
\nonumber\\
\langle\mathcal{P}^{\chi_{c0}}(^{3}P_{0}^{[1]})\rangle
&=&\frac{1}{2J+1}\langle\mathcal{P}^{\chi_{cJ}}(^{3}P_{J}^{[1]})\rangle,
\nonumber\\
\langle\mathcal{P}^{\chi_{c0}}(^{3}S_{1}^{[8]},^{3}D_{1}^{[8]})\rangle
&=&-\frac{2}{3}\langle\mathcal{P}^{\chi_{c1}}(^{3}S_{1}^{[8]},^{3}D_{1}^{[8]})\rangle
=2\langle\mathcal{P}^{\chi_{c2}}(^{3}S_{1}^{[8]},^{3}D_{1}^{[8]})\rangle.
\end{eqnarray}

The SDCs $F_{ij}(n)$ and $G_{ij}(n)$ may be obtained perturbatively by matching
the QCD and NRQCD calculations via the condition
\begin{equation}\label{mathching}
\sigma(c\bar{c})|_{\mathrm{pert\;QCD}}
=\sum_{n}\frac{F_{n}(\Lambda)}{m_c^{d_{\mathcal{O}(n)}-4}}
\langle0|\mathcal{O}_{n}^{c\bar{c}}(\Lambda)|0\rangle|_{\mathrm{pert\;NRQCD}}
+\sum_{n}\frac{G_{n}(\Lambda)}{m_c^{d_{\mathcal{P}(n)}-4}}
\langle0|\mathcal{P}_{n}^{c\bar{c}}(\Lambda)|0\rangle|_{\mathrm{pert\;NRQCD}}.
\end{equation}
The left-hand side of Eq.~(\ref{mathching}) may be computed directly using the
spinor projection method developed in Ref.~\cite{Kuhn:1979bb}, by which the
product of Dirac spinors $v(P/2-q)\bar{u}(P/2+q)$ is projected onto the
considered $^{2S+1}L_J$ state in a Lorentz-covariant form.
In an arbitrary reference frame, the four-momenta $P/2+q$ and $P/2-q$ of the
heavy quark and antiquark may be related to those in the quarkonium rest frame
as
\begin{equation}
\frac{P}{2}+q=L\left(\frac{P_r}{2}+\boldsymbol{q}\right),
\qquad
\frac{P}{2}-q=L\left(\frac{P_r}{2}-\boldsymbol{q}\right),
\end{equation}
where $P^{\mu}_r=(2E_q,\boldsymbol{0})$, $E_q=\sqrt{m_c^2+\boldsymbol{q}^2}$,
$2\boldsymbol{q}$ is the relative three-momentum between the two quarks in the
quarkonium rest frame, and $L^{\mu}_{\phantom{\mu}\nu}$ is the Lorentz
transformation matrix for the boost from the quarkonium rest frame to the
considered reference frame.
To all orders in $v^2$, the projectors onto the spin-singlet $(S=0)$ and
spin-triplet $(S=1)$ states in the quarkonium rest frame read
\cite{Bodwin:2002hg}
\begin{eqnarray}
\sum_{\lambda_{1},\lambda_{2}}v(-\boldsymbol{q},\lambda_{2})
\bar{u}(\boldsymbol{q},\lambda_{1})
\left\langle\frac{1}{2},\lambda_{1};\frac{1}{2},\lambda_{2}|0,0\right\rangle
&=&\frac{E_q+m_c}{\sqrt{2}}
\left(1-\frac{\boldsymbol{\alpha}\cdot\boldsymbol{q}}{E_q+m_c}\right)
\nonumber\\
&&{}\times
\gamma^{5}\frac{1+\gamma^{0}}{2}
\left(1+\frac{\boldsymbol{\alpha}\cdot\boldsymbol{q}}{E_q+m_c}\right)\gamma^{0},
\nonumber\\
\sum_{\lambda_{1},\lambda_{2}}v(-\boldsymbol{q},\lambda_{2})
\bar{u}(\boldsymbol{q},\lambda_{1})
\left\langle\frac{1}{2},\lambda_{1};\frac{1}{2},\lambda_{2}|1,
\boldsymbol{\epsilon}\right\rangle
&=&\frac{E_q+m_c}{\sqrt{2}}
\left(1-\frac{\boldsymbol{\alpha}\cdot\boldsymbol{q}}{E_q+m_c}\right)
\nonumber\\
&&{}\times
\boldsymbol{\alpha}\cdot\boldsymbol{\epsilon}\frac{1+\gamma^{0}}{2}
\left(1+\frac{\boldsymbol{\alpha}\cdot\boldsymbol{q}}{E_q+m_c}\right)\gamma^{0}.
\end{eqnarray}
In an arbitrary reference frame, they become
\begin{eqnarray}
\sum_{\lambda_{1},\lambda_{2}}v(-q,\lambda_{2})\bar{u}(q,\lambda_{1})
\left\langle\frac{1}{2},\lambda_{1};\frac{1}{2},\lambda_{2}|0,0\right\rangle
&=&\frac{-1}{2\sqrt{2}(E_q+m_c)}
\left(\frac{\slashed{P}}{2}-\slashed{q}-m_c\right)
\nonumber\\
&&{}\times
\gamma^{5}\frac{\slashed{P}+2E_q}{2E_q}
\left(\frac{\slashed{P}}{2}+\slashed{q}+m_c\right),
\nonumber\\
\sum_{\lambda_{1},\lambda_{2}}v(-q,\lambda_{2})\bar{u}(q,\lambda_{1})
\left\langle\frac{1}{2},\lambda_{1};\frac{1}{2},\lambda_{2}|1,\epsilon
\right\rangle
&=&\frac{-1}{2\sqrt{2}(E_q+m_c)}
\left(\frac{\slashed{P}}{2}-\slashed{q}-m_c\right)
\nonumber\\
&&{}\times
\slashed{\epsilon}\frac{\slashed{P}+2E_q}{2E_q}
\left(\frac{\slashed{P}}{2}+\slashed{q}+m_c\right).
\end{eqnarray}
Note that the normalization of the Dirac spinors is $\bar{u}u=-\bar{v}v=m_c^2$.
With the help of the spinor projection method, the partonic scattering
amplitude $M(ij\to c\bar{c}(n)+X)$ may then be expanded in the relative
momentum $q$.
To this end, we write
\begin{equation}
M(ij\to c\bar{c}(n)+X)=\sqrt{\frac{m_c}{E_q}}A(q),
\label{eq:exp}
\end{equation}
where the factor $\sqrt{m_c/E_q}$ stems from the relativistic normalization of
the $c\bar{c}(n)$ state and
\begin{eqnarray}
A(q)&=&\sum_{\lambda_{1},\lambda_{2}}\sum_{k,l}
\left\langle\frac{1}{2},\lambda_{1};\frac{1}{2},\lambda_{2}|S,S_z\right\rangle
\langle3,k;\bar{3},l|1(8,a)\rangle
\nonumber\\
&&{}\times\mathcal{A}\left(ij\to c_{\lambda_1,k}\left(\frac{P}{2}+q\right)
\bar{c}_{\lambda_2,l}\left(\frac{P}{2}-q\right)+X\right).
\label{eq:A}
\end{eqnarray}
Here, $\langle3,k;\bar{3},l|1\rangle=\delta_{kl}/\sqrt{N_c}$ and
$\langle3,k;\bar{3},l|8,a\rangle=\sqrt{2}\,T^{a}_{kl}$ are the color-SU(3)
Clebsch--Gordan coefficients for the $c\bar{c}(n)$ pair projected onto CS and CO
states, respectively, and
$\mathcal{A}(ij\to c_{\lambda_1,k}(P/2+q)\bar{c}_{\lambda_2,l}(P/2-q)+X)$ is
the standard Feynman amplitude.
Defining
\begin{equation}
A_{\alpha_1\cdots\alpha_N}(0)
=\left.\frac{\partial^NA(q)}{\partial q^{\alpha_1}\cdots\partial q^{\alpha_N}}
\right|_{q=0},
\end{equation}
we may write the expansion of Eq.~(\ref{eq:A}) in $q$ as
\begin{equation}
A(q)=A(0)+q^{\alpha_1}A_{\alpha_1}(0)
+\frac{1}{2}q^{\alpha_1}q^{\alpha_2}A_{\alpha_1\alpha_2}(0)
+\frac{1}{6}q^{\alpha_1}q^{\alpha_2}q^{\alpha_3}A_{\alpha_1\alpha_2\alpha_3}(0)
+\cdots.
\label{eq:Aq}
\end{equation}
For $S$- and $D$-wave states, only the terms with even powers in $q$
contribute, while for $P$-wave states it is the other way around.
To calculate the relativistic corrections for the production of $S$- and
$P$-wave states, we need to decompose the higher-rank tensor products of $q$
factors in Eq.~(\ref{eq:Aq}) into their irreducible representations and to
retain the $L=S$ and $L=P$ terms, respectively.
Through order $v^2$, we thus obtain,
for $n={}^3\!S_{1}^{[1]},{}^3\!S_{1}^{[8]},{}^1\!S_{0}^{[8]}$,
\begin{equation}
M(i j\to c\bar{c}(n)+X)=\sqrt{\frac{m_c}{E_q}}
\left[A(0)+\frac{|\boldsymbol{q}|^2}{6} \Pi^{\alpha_1\alpha_2}A_{\alpha_1\alpha_2}(0)
\right],
\end{equation}
and, for $n={}^3\!P_{J}^{[1]},{}^3\!P_{J}^{[8]}$,
\begin{equation}
M(i j\to c\bar{c}(n)+X)=\sqrt{\frac{m_c}{E_q}}q^{\alpha_1}\left[A_{\alpha_1}(0)
-\frac{|\boldsymbol{q}|^2}{30}
\Pi_{\alpha_1}^{\phantom{\alpha_1}\alpha_2\alpha_3\alpha_4}A_{\alpha_2\alpha_3\alpha_4}(0)\right],
\end{equation}
where
$\Pi^{\alpha_1\alpha_2}
=-g^{\alpha_1\alpha_2}+P^{\alpha_1}P^{\alpha_2}/(4E_q^2)$ and
$\Pi^{\alpha_1\alpha_2\alpha_3\alpha_4}=\Pi^{\alpha_1\alpha_2}\Pi^{\alpha_3\alpha_4}
+\Pi^{\alpha_1\alpha_3}\Pi^{\alpha_2\alpha_4}+\Pi^{\alpha_2\alpha_3}\Pi^{\alpha_1\alpha_4}$.
In the $D$-wave case, we only need the amplitude at LO in $v^2$, which is
\begin{equation}
M(i j\to c\bar{c}(^3D_J^{[8]})+X)
=\sqrt{\frac{m_c}{E_q}}\,\frac{1}{2}q^{\alpha_1}q^{\alpha_2}A_{\alpha_1\alpha_2}(0).
\end{equation}

We are now in a position to perform the matching between the calculations in
NRQCD and full QCD.
We thus obtain
\begin{eqnarray}
\frac{F_{ij}(n)}{m_c^{d_{\mathcal{O}(n)}-4}}&=&\frac{1}{2s}
\int d\mathrm{LIPS}|\overline{M}_{ij}(n)|^2,
\nonumber\\
\frac{G_{ij}(n)}{m_c^{d_{\mathcal{P}(n)}-4}}&=&\frac{1}{2s}\int d\mathrm{LIPS}
\left(K|\overline{M}_{ij}(n)|^2+|\overline{N}_{ij}(n)|^2\right),
\label{eq:fg1}
\end{eqnarray}
where $\sqrt{s}$ is the invariant mass of the incoming partons and
\begin{eqnarray}
|\overline{M}_{ij}(n)|^2&=&
\overline{\sum_{L_z}}|A(0)|^2\bigg|_{\boldsymbol{q}^2=0},
\nonumber\\
|\overline{N}_{ij}(n)|^2&=&
\overline{\sum_{L_z}}
\left\{\frac{\partial}{\partial\boldsymbol{q}^2}
\left[\frac{m_c}{E_q}|A(0)|^2\right]
+\frac{1}{3}\Pi^{\alpha_1\alpha_2}\re[A^*(0)A_{\alpha_1\alpha_2}(0)]\right\}
\bigg|_{\boldsymbol{q}^2=0},
\label{eq:mnd}
\end{eqnarray}
for $n={}^3\!S_{1}^{[1]},{}^3\!S_{1}^{[8]},{}^1\!S_{0}^{[8]}$;
\begin{eqnarray}
|\overline{M}_{ij}(n)|^2&=&\overline{\sum_{S_z,L_z,J_z}}
|\langle1,L_z;1,S_z|J,J_z\rangle
\epsilon^{\ast\alpha}_{L_z}A_{\alpha}(0)|^2\bigg|_{\boldsymbol{q}^2=0},
\nonumber\\
|\overline{N}_{ij}(n)|^2&=&\left\{
\overline{\sum_{S_z,L_z,J_z}}\frac{\partial}{\partial\boldsymbol{q}^2}
\left[\frac{m_c}{E_q}|\langle1,L_z;1,S_z|J,J_z\rangle
\epsilon^{\ast\alpha}_{L_z}A_{\alpha}(0)|^2\right]
\right.\nonumber\\
&&{}-\frac{1}{15}\overline{\sum_{S_z,L_z,J_z}}\langle1,L_z;1,S_z|J,J_z\rangle
\overline{\sum_{S^{\prime}_z,L^{\prime}_z,J^{\prime}_z}}
\langle1,L^{\prime}_z;1,S^{\prime}_z|J,J^{\prime}_z\rangle
\nonumber\\
&&{}\times\left.\vphantom{\overline{\sum_{S_z,L_z,J_z}}}
\Pi_\alpha^{\phantom{\alpha}\alpha_1\alpha_2\alpha_3}
\re[\epsilon^{\ast\alpha}_{L_z}\epsilon^{\beta}_{L^{\prime}_z}
A_{\beta}^*(0)A_{\alpha_1\alpha_2\alpha_3}(0)]\right\}\bigg|_{\boldsymbol{q}^2=0},
\qquad
\end{eqnarray}
for $n={}^3\!P_{J}^{[1]},{}^3\!P_{J}^{[8]}$; and
\begin{equation}
|\overline{N}_{ij}(n)|^2=\overline{\sum_{S_z,L_z,J_z}}
\langle2,L_z;1,S_z|1,J_z\rangle
\re[\epsilon^{\ast\alpha\beta}_{L_z}A^*(0)A_{\alpha\beta}(0)]
\bigg|_{\boldsymbol{q}^2=0},
\label{eq:mnm}
\end{equation}
for the ${}^3\!S_1^{[8]}$-${}^3\!D_1^{[8]}$ mixing term.
Here, $\epsilon_{L_z}^{\alpha}$ ($\epsilon_{L_z}^{\alpha\beta}$) is the
polarization four-vector (four-tensor) for $L=P$ ($D$), the symbol $\sum$
also implies the summation over the polarizations of the other external
partons, and the bar implies the average over the spins and colors of the
incoming partons and those of the $c\bar{c}(n)$ state.
The factor $K$ in Eq.~(\ref{eq:fg1}) contains the $v^2$ corrections to the
phase space and depends on the kinematic variables that we are interested in.
In the cases under consideration here, we have $K=-4/(s-4m_c^2)$. 
We generate the Feynman diagrams using the FeynArts package
\cite{Kublbeck:1990xc} and compute the amplitude squares using the FeynCalc
package \cite{Mertig:1990an}.
In the Appendix, we present our results for $|\overline{N}_{ij}(n)|^2$.


We reproduce the well-known results for $|\overline{M}_{ij}(n)|^2$, which were
called  $|\mathcal{A}|^2$ in Ref.~\cite{Cho:1995vh} and
$|\mathcal{M}^{\prime}|^2$ in Ref.~\cite{Ko:1996xw}.
We may also compare some of our results for $|\overline{N}_{ij}(n)|^2$ with the
literature.
The $v^2$ corrections to direct $J/\psi$ photoproduction were first studied in
Ref.~\cite{Jung:1993cd} within the relativistic quark model, which accounts for
the CS contributions.
We reproduce Eq.~(23) in Ref.~\cite{Jung:1993cd} if we do not expand
$E_q$, but set $E_q=m_{J/\psi}/2$.
The relativistic corrections to direct $J/\psi$ hadroproduction were considered
in Ref.~\cite{Xu:2012am} within NRQCD.
We find agreement with Ref.~\cite{Xu:2012am}, except for some typographical
errors in Eq.~(A4) therein, which corresponds to
$|\overline{N}_{gg}({}^3\!S_1^{[8]})|^2$ in our notation.
Furthermore, the ${}^3\!S_1^{[8]}$-${}^3\!D_1^{[8]}$ mixing contribution was
not considered there.
Apart from correcting the misprints in Eq.~(A4) of Ref.~\cite{Xu:2012am}, our
paper reaches beyond the available literature by studying the CO contributions
to direct $J/\psi$ photoproduction, the ${}^3\!S_1^{[8]}$-${}^3\!D_1^{[8]}$
mixing contribution to direct $J/\psi$ hadroproduction, and the feed-down
contributions to $J/\psi$ hadroproduction.

\section{Phenomenological results}
\label{sec:three}

\begin{figure}
\begin{center}
\begin{tabular}{ccc}
\includegraphics[scale=0.50]{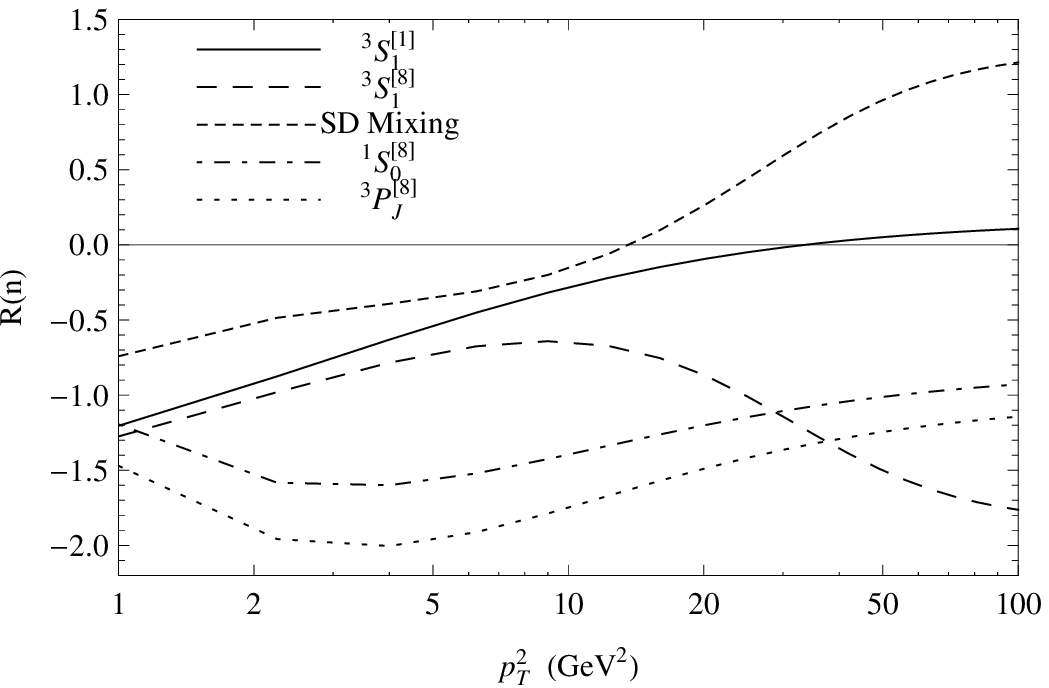} &
\includegraphics[scale=0.50]{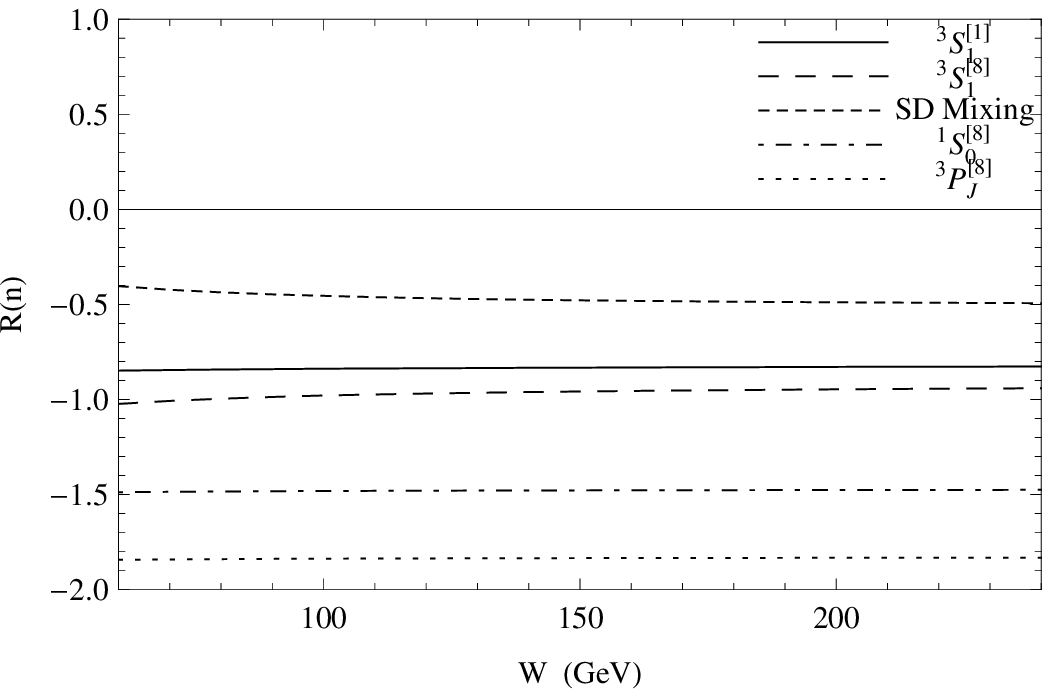} &
\includegraphics[scale=0.50]{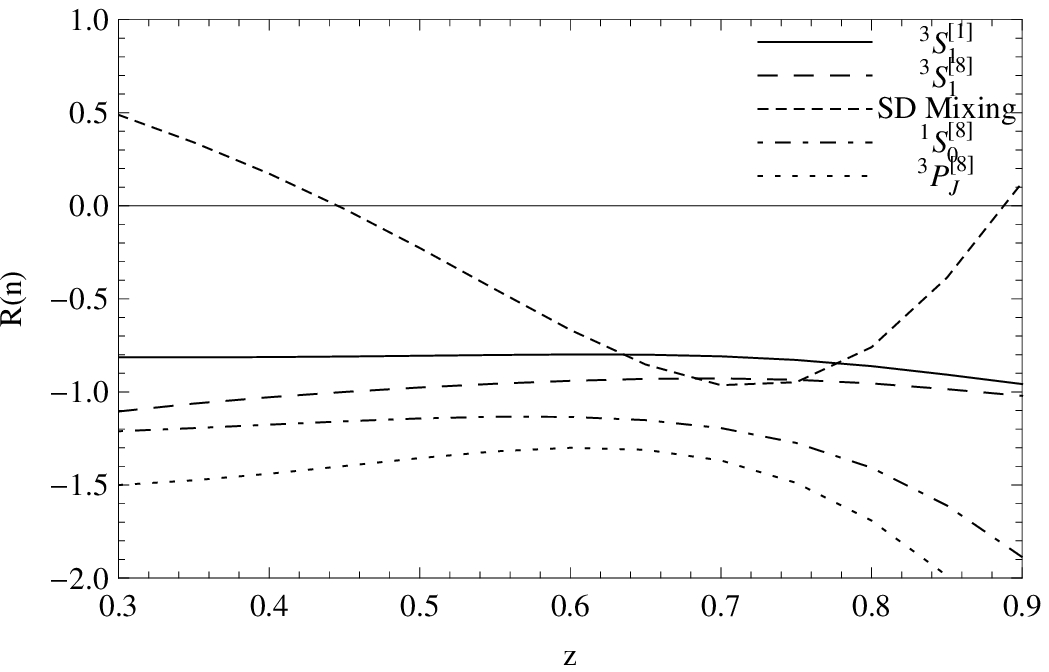} \\
(a) & (b) & (c) \\
\end{tabular}
\caption{Ratios $R(n)$ for $J/\psi$ direct photoproduction under HERA~II
kinematic conditions as functions of (a) $p_T^2$, (b) $W$, and (c) $z$.}
\label{fig:one}
\end{center}
\end{figure}

We are now in a position to investigate the phenomenological significance of
the $v^2$ corrections in prompt $J/\psi$ photoproduction and hadroproduction.
In our numerical analysis, we use $m_c=1.5$~GeV, $\alpha=1/137.036$, the LO
formula for $\alpha_s^{(n_f)}(\mu_r)$ with $n_f=4$ active quark flavors and
asymptotic scale parameter $\Lambda_\mathrm{QCD}^{(4)}=215$~MeV
\cite{Pumplin:2002vw}, the CTEQ6L1 set for proton PDFs \cite{Pumplin:2002vw},
the photon flux function given in Eq.~(5) of Ref.~\cite{Kniehl:1996we} with
$Q_{\rm max}^2=2.5~\mathrm{GeV}^2$ \cite{Aaron:2010gz}, and the choice
$\mu_r=\mu_f=\sqrt{p_T^2+4m_c^2}$ for the renormalization and factorization
scales.
According to Eqs.~(\ref{xs}) and (\ref{eq:fg}), the hadronic cross sections of
the direct photoproduction and hadroproduction of the charmonia
$H=J/\psi,\chi_{cJ},\psi^\prime$, differential in some observable $x$, may be
generically written as
\begin{equation}
\frac{d\sigma}{dx}=\sum_n\left(
\frac{dF(n)}{dx}\,\frac{\langle\mathcal{O}^{H}(n)\rangle}
{m_c^{d_{\mathcal{O}(n)}-4}}
+\frac{dG(n)}{dx}\,\frac{\langle\mathcal{P}^{H}(n)\rangle}
{m_c^{d_{\mathcal{P}(n)}-4}}\right),
\end{equation}
where it is understood that $dF(n)/dx=0$ if $n$ stands for
${}^3\!S_1^{[8]}$-${}^3\!D_1^{[8]}$ mixing.
For all other channels $n$, the relative $v^2$ corrections are
$R(n)\langle\mathcal{P}^{H}(n)\rangle/(m_c^2\langle\mathcal{O}^{H}(n)\rangle)$,
where $R(n)=dG(n)/dx/(dF(n)/dx)$ is a dimensionless ratio.
To standardize the numerical discussion, we also introduce the quantity
$R(n)=dG(n)/dx/(dF({}^3\!S_1^{[8]})/dx)$ for the case when $n$ stands for
${}^3\!S_1^{[8]}$-${}^3\!D_1^{[8]}$ mixing.
According to the velocity scaling rules \cite{Lepage:1992tx}, the weight
$\langle\mathcal{P}^{H}(n)\rangle/(m_c^2\langle\mathcal{O}^{H}(n)\rangle)$ of
$R(n)$ is of order $v^2$, which is approximately 0.23 for charmonium
\cite{Bodwin:2007fz,Guo:2011tz}.
For definiteness, we ignore these weights in the following and concentrate on
the SDC ratios $R(n)$ instead.
Furthermore, we limit ourselves to the direct production of $J/\psi$ mesons and
their production via the feed down from the $\chi_{cJ}$ mesons.
In the latter case, the branching fractions $B(\chi_{cJ}\to J/\psi+X)$ drop out
in the ratios $R(n)$.
However, there remains a kinematic effect on the transverse momentum.
Since $p_T\equiv p_T^{J/\psi}\gg M_{\chi_{cJ}}-M_{J/\psi}$, we may approximate
$p_T^{J/\psi}=p_T^{\chi_{cJ}}M_{J/\psi}/M_{\chi_{cJ}}$, with
$M_{J/\psi}=3.097~\mathrm{GeV}$, $M_{\chi_{c0}}=3.415~\mathrm{GeV}$,
$M_{\chi_{c1}}=3.511~\mathrm{GeV}$, and $M_{\chi_{c2}}=3.556~\mathrm{GeV}$
\cite{Beringer:1900zz}.

\begin{figure}
\begin{center}
\begin{tabular}{cc}
\includegraphics[scale=0.75]{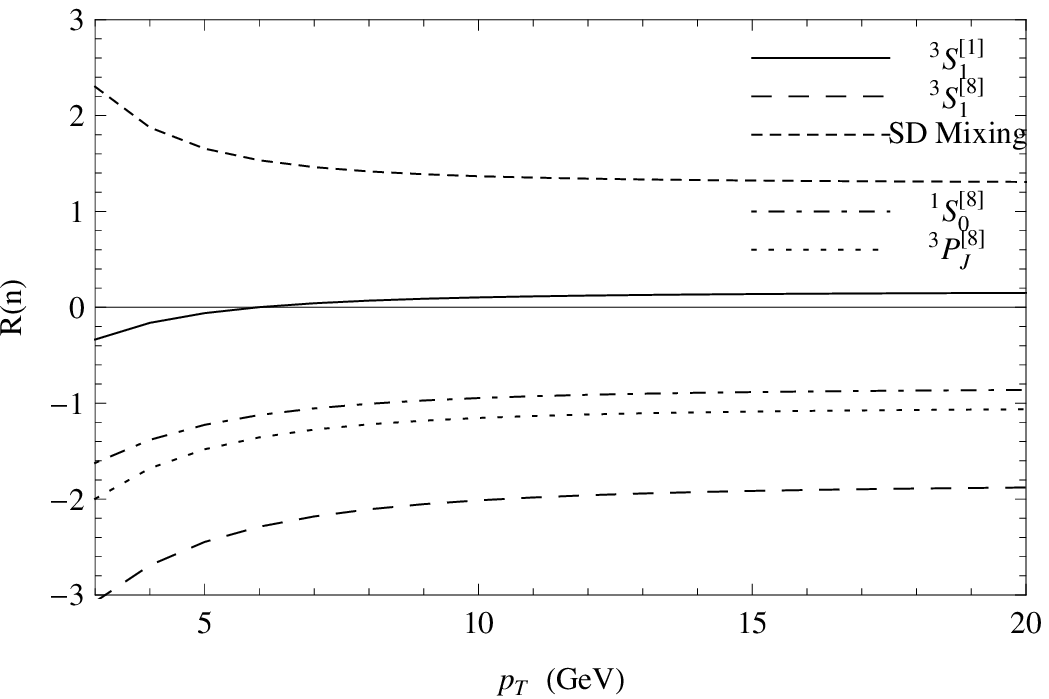} &
\includegraphics[scale=0.75]{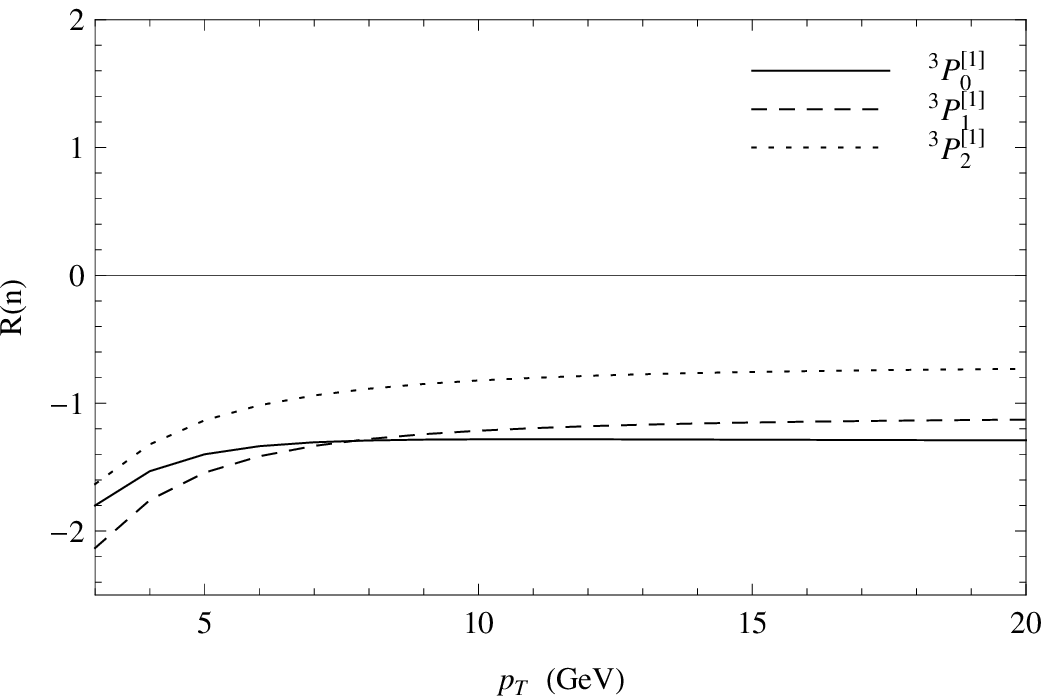} \\
(a) & (b) \\
\end{tabular}
\caption{Ratios $R(n)$ for $J/\psi$ hadroproduction (a) in the direct mode
and (b) via the feed down from $\chi_{cJ}$ mesons under Tevatron Run~I
kinematic conditions as functions of $p_T$.}
\label{fig:two}
\end{center}
\end{figure}

\begin{figure}
\begin{center}
\begin{tabular}{cc}
\includegraphics[scale=0.75]{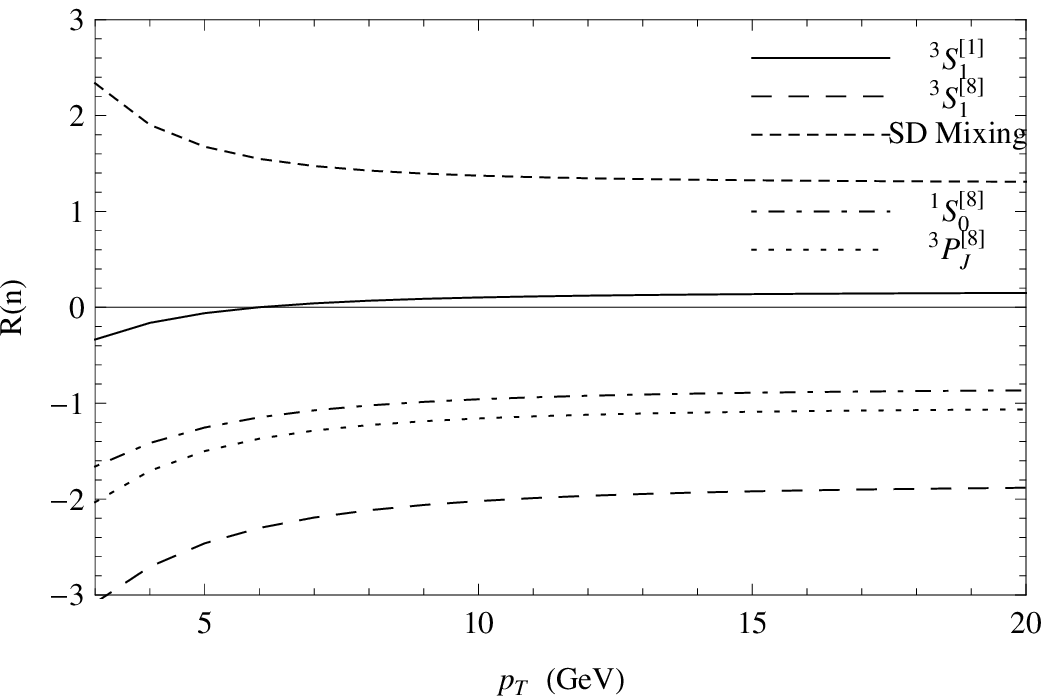} &
\includegraphics[scale=0.75]{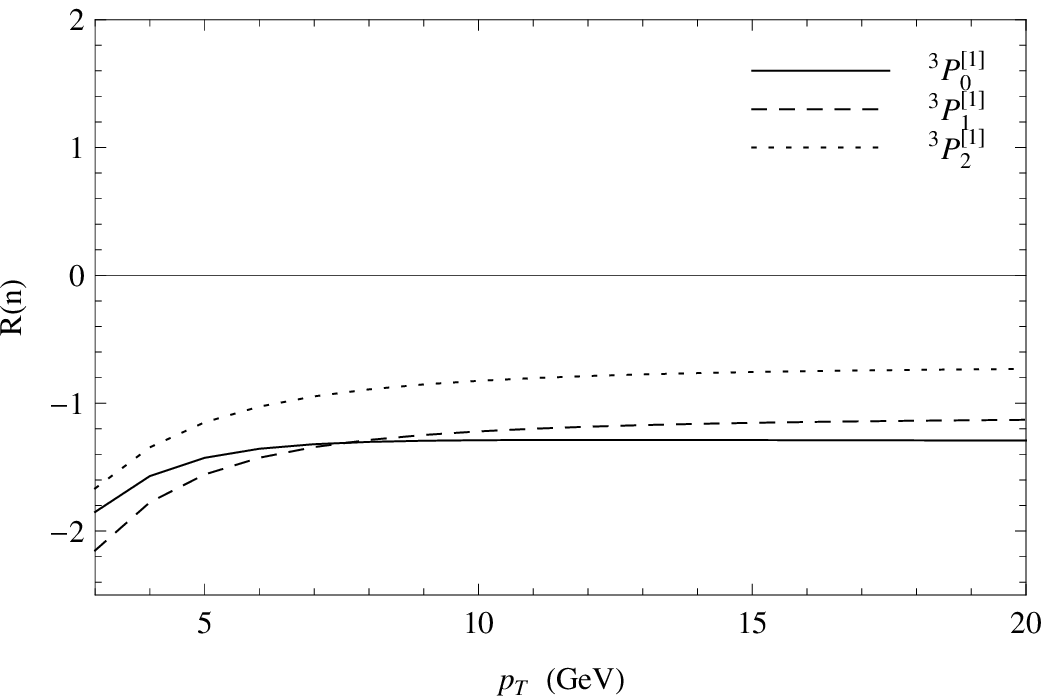} \\
(a) & (b) \\
\end{tabular}
\caption{Ratios $R(n)$ for $J/\psi$ hadroproduction (a) in the direct mode
and (b) via the feed down from $\chi_{cJ}$ mesons under LHCb kinematic
conditions as functions of $p_T$.}
\label{fig:three}
\end{center}
\end{figure}

We consider three typical experimental environments, namely, Run~II at HERA,
Run~I at the Tevatron, and the LHCb setup at the LHC.
At HERA~II, the cross section of prompt $J/\psi$ photoproduction was measured
at center-of-mass energy $\sqrt{S}=319~\mathrm{GeV}$ differential in $p_T^2$,
$W=\sqrt{(p_p+p_{\gamma})^2}$, and
$z=p_{J/\psi}\cdot p_{p}/p_{\gamma}\cdot p_{p}$ \cite{Aaron:2010gz}, where
$p_{p}$, $p_{\gamma}$, and $p_{J/\psi}$ are the four-momenta of the proton,
photon, and $J/\psi$ meson, respectively, imposing in turn two of the
acceptance cuts $1~\mathrm{GeV}^2<p_T^2$, $60~\mathrm{GeV}<W<240~\mathrm{GeV}$,
and $0.3<z<0.9$.
In Figs.~\ref{fig:one}(a)--(c), the $p_T^2$, $W$, and $z$ distributions of
$R(n)$ are shown for
$n={}^3\!S_1^{[1]},{}^3\!S_1^{[8]},{}^1\!S_0^{[8]},{}^3\!P_J^{[8]}$, and
${}^3\!S_1^{[8]}$-${}^3\!D_1^{[8]}$ mixing.
We recall that $n={}^3\!P_J^{[1]}$ is prohibited at this order.
For simplicity, we ignore resolved photoproduction, the SDCs of which also
contribute to hadroproduction to be studied below.
In Run~I at the Tevatron, the $p_T$ distribution of prompt $J/\psi$
hadroproduction was measured at $\sqrt{S}=1.8~\mathrm{TeV}$ in the
pseudorapidity range $|\eta|<0.6$ \cite{Abe:1997jz}, and it was measured by
LHCb at $\sqrt{S}=7~\mathrm{TeV}$ in the rapidity range $2.0<y<4.5$
\cite{Aaij:2011jh}.
In the latter two cases, we exclude the region $p_T<3~\mathrm{GeV}$, where the
application of fixed-order perturbation theory is problematic.
In Figs.~\ref{fig:two} and \ref{fig:three}, the $p_T$ distributions of $R(n)$
are shown for the Tevatron and the LHC, respectively.
In each figure, part~(a) refers to
$n={}^3\!S_1^{[1]},{}^3\!S_1^{[8]},{}^1\!S_0^{[8]},{}^3\!P_J^{[8]}$, and
${}^3\!S_1^{[8]}$-${}^3\!D_1^{[8]}$ mixing in direct $J/\psi$ production, and
part~(b) refers to $n={}^3\!P_J^{[1]}$ in the feed down from the respective
$\chi_{cJ}$ meson.
In part~(b), we do not show $R(n)$ for $n={}^3\!S_1^{[8]}$ and
${}^3\!S_1^{[8]}$-${}^3\!D_1^{[8]}$ mixing because these results emerge from
their counterparts in part~(a) just by rescaling the $p_T$ axis as explained
above.
In a similar way, part~(a) carries over to the feed down from the $\psi^\prime$
meson.

We observe from Figs.~\ref{fig:one}--\ref{fig:three} that the ratios $R(n)$
may be throughout positive or negative or change sign as $x$ is varied and
typically have a magnitude of order unity or larger.
In fact, $R({}^3\!S_1^{[8]})$ for hadroproduction reaches the value $-3$ at
$p_T=3~\mathrm{GeV}$, as may be seen in Figs.~\ref{fig:two} and
\ref{fig:three}.
An exception to this rule is $R({}^3\!S_1^{[1]})$ for hadroproduction, which is
roughly one order of magnitude smaller.
Because of the above estimations of the weights
$\langle\mathcal{P}^{H}(n)\rangle/(m_c^2\langle\mathcal{O}^{H}(n)\rangle)$ and
$v^2$, $R(n)=2$ is likely to imply a $v^2$ correction of some 50\% in the
channel $n$.
However, it is unlikely that the inclusion of the $v^2$ corrections would turn
a given hadronic cross section negative, the more so as the $v^2$ corrections
in the various channels $n$ may be of either sign thus allowing for
cancellations.   
Comparing Figs.~\ref{fig:one}--\ref{fig:three}, we observe that the relative
importance and the $p_T$ dependencies of the various ratios $R(n)$ greatly
differ between photoproduction and hadroproduction.
If a fit is constrained to data for which the contributing ratios $R(n)$ are
approximately independent of $x$, as in the case of hadroproduction for
$p_T>7~\mathrm{GeV}$ say, then the inclusion of the $v^2$ corrections would not
improve the quality of the fit, but rather induce strong correlations between
$\langle\mathcal{O}^{H}(n)\rangle$ and $\langle\mathcal{P}^{H}(n)\rangle$ for
each $n$, both as for sign and magnitude.
To reduce these correlations or even enable independent determinations of the
LDMEs in the enlarged set, it is indispensable to reduce the low-$p_T$ cut in
hadroproduction, to $p_T>3~\mathrm{GeV}$ say, and to include photoproduction
data in the fit.
In a fit with such a low-$p_T$ cut in hadroproduction, the inclusion of the
$v^2$ corrections might enhance the goodness because the notorious lack of
turnover of the NLO NRQCD predictions in the low-$p_T$ region, which causes
the latter to overshoot the experimental data there, might be cured at least
to some extent by the fact that all the ratios $R(n)$ are amplified as $p_T$ is
decreased.
In conclusion, the addition of $v^2$ corrections is expected to have a
significant impact on state-of-the-art determinations of CO LDMEs through
global fits to prompt $J/\psi$ production data and might improve the goodness
of such fits.

\section{Conclusions}
\label{sec:four}

We performed a systematic study of the $v^2$ corrections to the yields of
prompt $J/\psi$ photoproduction and hadroproduction, providing the relevant
SDCs in analytic form.
Specifically, this includes the partonic subprocesses listed in
Eq.~(\ref{eq:sub}) in combination with the $c\bar{c}$ Fock states
$n={}^3\!S_1^{[1]},{}^3\!P_J^{[1]},{}^3\!S_1^{[8]},{}^1\!S_0^{[8]},
{}^3\!P_J^{[8]}$, and ${}^3\!S_1^{[8]}$-${}^3\!D_1^{[8]}$ mixing.
We compared our results with the literature as far as the latter goes.
We assessed the phenomenological significance of the $v^2$ corrections in the
various channels by studying their ratios with respect to the corresponding LO
results.
Assuming the relevant LDMEs $\langle\mathcal{O}^H(n)\rangle$ and
$\langle\mathcal{P}^H(n)\rangle$ to obey the hierarchy predicted by the
velocity scaling rules \cite{Lepage:1992tx}, we found that $v^2$ corrections of
up to 50\% are realistic.
We thus conclude that it is indispensable to include $v^2$ corrections in
determinations of CO LDMEs by global data fits, the more so as the $v^2$
corrections in the various channels greatly differ between photoproduction and
hadroproduction.

\section*{Acknowledgments}

Z.-G.H. would like to thank Yu-Jie Zhang for a communication about the results
in Ref.~\cite{Xu:2012am}.
This work was supported in part by the German Federal Ministry for Education
and Research BMBF through Grant No.~05H12GUE.

\section{Appendix}

In this appendix, we present analytic expressions for the nonvanishing SDCs
$|\overline{N}_{ij}(n)|^2$ appropriate for prompt $J/\psi$ photoporduction and
hadroproduction evaluated according to Eqs.~(\ref{eq:mnd})--(\ref{eq:mnm}).
For the partonic subprocess $i(p_1)+j(p_2)\to c\bar{c}(P)+X$, the Mandelstam
variables are defined as $s=(p_1+p_2)^2$, $t=(p_1-P)^2|_{\boldsymbol{q}=0}$, and
$u=(p_2-P)^2|_{\boldsymbol{q}=0}$ and satisfy $s+t+u=4m_c^2$.
Note that our results for hadroproduction can also be applied to resolved
photoproduction.

\subsection{Photoproduction}

$g+\gamma\to c\bar{c}({}^3\!S_1^{[1]})+g$:


\begin{eqnarray}
&&|\overline{N}_{g\gamma}({}^3\!S_1^{[1]})|^2
=\frac{8192 \pi ^3 \alpha  \alpha_s^2} {729 m_c
\left(s-4 m_c^2\right)^3 \left(4 m_c^2-t\right)^3 (s+t)^3}\nonumber\\
&&\times\Big{[}2048 m_c^{10} \left(3 s^2+2 s t+3 t^2\right)-256 m_c^8
\left(5 s^3-8 s^2 t+4 s t^2+5 t^3\right)-64 m_c^6 (15 s^4+68 s^3 t\nonumber\\
&&+62 s^2 t^2+32 s t^3+15 t^4)+16 m_c^4 \left(21 s^5+87 s^4 t+130 s^3 t^2
+106 s^2 t^3+51 s t^4+21 t^5\right)\nonumber\\
&&-4 m_c^2 \left(7 s^6+30 s^5 t+59 s^4 t^2+64 s^3 t^3+47 s^2 t^4+18 s t^5+7 t^6\right)-s t (s+t)
\left(s^2+s t+t^2\right)^2\Big{]}\nonumber\\
\end{eqnarray}

$g+\gamma\to c\bar{c}({}^3\!S_1^{[8]})+g$:

\begin{equation}
|\overline{N}_{g\gamma}({}^3\!S_1^{[8]})|^2
=\frac{15}{8}|\overline{N}_{g\gamma}(^3S_1^{[1]})|^2
\end{equation}

$g+\gamma\to c\bar{c}({}^1\!S_0^{[8]})+g$:

\begin{eqnarray}
&&|\overline{N}_{g\gamma}({}^1\!S_0^{[8]})|^2
=\frac{128 \pi ^3 \alpha  \alpha_s^2 s t}{9 m_c^3 \left(4 m_c^2
-s\right)^3\left(4 m_c^2-t\right)^3 (s+t)^3 \left(-4 m_c^2+s+t\right)}\nonumber\\
&&\times\Big{[}131072 m_c^{14}-4096 m_c^{12} (23 s+11 t)+1024 m_c^{10} \left(33 s^2
+34 s t-3 t^2\right)-256 m_c^8 (s+t) \nonumber\\
&&\left(7 s^2-11 s t-29 t^2\right)-64 m_c^6 \left(13 s^4+94 s^3 t+170 s^2 t^2+142 s t^3
+37 t^4\right)+16 m_c^4 (23 s^5\nonumber\\
&&+112 s^4 t+214 s^3 t^2+226 s^2 t^3+112 s t^4+23 t^5)-4 m_c^2 (9 s^6+50 s^5 t
+116 s^4 t^2+138 s^3 t^3\nonumber\\
&&+104 s^2 t^4+38 s t^5+9 t^6)+5 s t (s+t) \left(s^2+s t+t^2\right)^2\Big{]}
\end{eqnarray}

$g+\gamma\to c\bar{c}({}^3\!P_J^{[8]})+g$:


\begin{eqnarray}
&&|\overline{N}_{g\gamma}({}^3\!P_J^{[8]})|^2
=\frac{128 \pi ^3 \alpha  \alpha_s^2}{15 m_c^5
\left(s-4 m_c^2\right)^4 \left(t-4 m_c^2\right)^4 (s+t)^4
\left(-4m_c^2+s+t\right)}\nonumber\\
&&\times\Big{[}3145728 m_c^{18} (s+t) \left(s^2+8 s t+t^2\right)
-65536 m_c^{16} (18 s^4+113 s^3 t+370 s^2 t^2-27 s t^3\nonumber\\
&&+18 t^4)-32768 m_c^{14} \left(39 s^5+318 s^4 t+459 s^3 t^2+929 s^2 t^3
+548 s t^4+39 t^5\right)+16384 m_c^{12} \nonumber\\
&&\left(63 s^6+519 s^5 t+1200 s^4 t^2+2043 s^3 t^3+1805 s^2 t^4
+629 s t^5+63 t^6\right)-2048 m_c^{10} (144 s^7 \nonumber\\
&&+1231s^6 t+3673 s^5 t^2+7074 s^4 t^3+7874 s^3 t^4+4843 s^2 t^5
+1301 s t^6 +144 t^7)+512 m_c^8 \nonumber\\
&&\left(75 s^8+662 s^7 t+2626 s^6 t^2+5981 s^5 t^3+7990 s^4 t^4+6371 s^3 t^5
+3206 s^2 t^6+642 s t^7+75 t^8\right)\nonumber\\
&&-128 m_c^6 (15 s^9+162 s^8 t+964 s^7 t^2+2725 s^6 t^3
+4364 s^5 t^4+4034 s^4 t^5+2635 s^3 t^6+1114 s^2 t^7\nonumber\\
&&+132 s t^8+15 t^9)+16 m_c^4 s t (27 s^8+502 s^7 t
+1968 s^6 t^2+3774 s^5t^3+4338 s^4 t^4+3114 s^3 t^5\nonumber\\
&&+1668 s^2 t^6+502 s t^7+27 t^8)-16 m_c^2 s^2 t^2 (s+t)
(25 s^6+118 s^5 t+259 s^4 t^2+310 s^3 t^3\nonumber\\
&&+244 s^2 t^4+103 s t^5+25 t^6)+31 s^3 t^3 (s+t)^2
\left(s^2+s t+t^2\right)^2\Big{]}
\end{eqnarray}

$g+\gamma\to c\bar{c}({}^3\!S_1^{[8]},{}^3\!D_1^{[8]})+g$:

\begin{eqnarray}
&&|\overline{N}_{g\gamma}({}^3\!S_1^{[8]},{}^3\!D_1^{[8]})|^2
=\frac{2048 \sqrt{5/3}
\pi ^3 \alpha  \alpha_{s}^{2}}{81 m_c \left(s-4 m_c^2\right)^4
\left(t-4 m_c^2\right)^4 (s+t)^4} \Big{[}32768 m_c^{14} (s+t) (3 s^2\nonumber\\
&&+5 s t+3 t^2)-4096 m_c^{12} \left(s^2+5 s t+t^2\right)
\left(11 s^2+18 s t+11 t^2\right)-1024 m_c^{10} (s+t) (10 s^4\nonumber\\
&&-109 s^3 t-202 s^2 t^2-109 s t^3+10 t^4)+256 m_c^8
(36 s^6-27 s^5 t-395 s^4 t^2-640 s^3 t^3\nonumber\\
&&-395 s^2 t^4-27 s t^5+36 t^6)-64 m_c^6 (s+t)
(28 s^6-3 s^5 t-250 s^4 t^2-400 s^3 t^3-250 s^2 t^4\nonumber\\
&&-3 s t^5+28 t^6)+16 m_c^4 (7 s^8+4 s^7 t
-100 s^6 t^2-323 s^5 t^3-440 s^4t^4-323 s^3 t^5-100 s^2 t^6\nonumber\\
&&+4 s t^7+7 t^8)+4 m_c^2 s t (s+t)
\left(6 s^6+22 s^5 t+51 s^4 t^2+66 s^3 t^3+51 s^2 t^4+22 s t^5+6  t^6\right)\nonumber\\
&&-s^2 t^2 (s+t)^2 \left(s^2+s t+t^2\right)^2\Big{]}
\end{eqnarray}



$q(\bar{q})+\gamma\to c\bar{c}({}^3\!S_1^{[8]})+q(\bar{q})$:


\begin{eqnarray}
&&|\overline{N}_{q(\bar{q})\gamma}({}^3\!S_1^{[8]})|^2
=\frac{8 \pi ^3 \alpha  \alpha_s^2 e_q^2}
{27 m_c^5 s t \left(4 m_c^2-s\right)}\nonumber\\
&&\times\Big{[}640 m_c^6-160 m_c^4 (2 s+t)+4 m_c^2 \left(27 s^2+10 s t
+5 t^2\right)-11 s \left(s^2+t^2\right)\Big{]}
\end{eqnarray}

$q(\bar{q})+\gamma\to c\bar{c}({}^1\!S_0^{[8]})+q(\bar{q})$:


\begin{eqnarray}
|\overline{N}_{q(\bar{q})\gamma}({}^1\!S_0^{[8]})|^2
=\frac{256 \pi ^3 \alpha  \alpha_s^2 \left[m_c^2 \left(44 s^3
+92 s^2 t-4 s t^2+44 t^3\right)-5 s (s+t)\left(s^2+t^2\right)\right]}{81 m_c^3
 \left(s-4 m_c^2\right) (s+t)^3 \left(-4 m_c^2+s+t\right)}
\end{eqnarray}

$q(\bar{q})+\gamma\to c\bar{c}({}^3\!P_J^{[8]})+q(\bar{q})$:


\begin{eqnarray}
&&|\overline{N}_{q(\bar{q})\gamma}({}^3\!P_J^{[8]})|^2
=\frac{256 \pi ^3 \alpha  \alpha_s^2}
{135 m_c^5 \left(s-4 m_c^2\right) (s+t)^4 \left(-4 m_c^2+s+t\right)}\nonumber\\
&&\times\Big{[}512 m_c^6 \left(5 s^2+26 s t+25 t^2\right)+
64 m_c^4 \left(s^3-23 s^2 t-111 s t^2-19 t^3\right)+4 m_c^2 (s+t)\nonumber\\
&& (57 s^3+169 s^2 t-3 s t^2+61 t^3)-31 s (s+t)^2 \left(s^2+t^2\right)\Big{]}
\end{eqnarray}

$q(\bar{q})+\gamma\to c\bar{c}({}^3\!S_1^{[8]},{}^3\!D_1^{[8]})+q(\bar{q})$:

\begin{eqnarray}
&&|\overline{N}_{q(\bar{q})\gamma}({}^3\!S_1^{[8]},{}^3\!D_1^{[8]})|^2
=-\frac{16 \sqrt{5/3} \pi ^3 \alpha  \alpha_{s}^{2}
e_q^2 \left[32 m_c^4-8 m_c^2 (s+t)+s^2+t^2\right]}{9 m_c^5 s t}
\end{eqnarray}

\subsection{Hadroproduction}

$g+g\to c\bar{c}({}^3\!S_1^{[1]})+g$:

\begin{eqnarray}
|\overline{N}_{gg}({}^3\!S_1^{[1]})|^2=\frac{15\alpha_s}{128\alpha}
|\overline{N}_{g\gamma}({}^3\!S_1^{[1]})|^2
\end{eqnarray}


$g+g\to c\bar{c}({}^3\!P_0^{[1]})+g$:

\begin{eqnarray}
&&|\overline{N}_{gg}({}^3\!P_0^{[1]})|^2
=\frac{8 \pi ^3 \alpha_{s}^{3}}
{45 m_c^5  s t \left(4 m_c^2-s\right)^5
\left(4 m_c^2-t\right)^5 (s+t)^5 \left(-4 m_c^2+s+t\right)}\nonumber\\
&&\times\Big{[}251658240 (s+t)^3 \left(11 s^2+30 t s+11 t^2\right)
m_c^{24}-4194304 (s+t)^2 (5 s+3 t) \big{(}231 s^3+847 t s^2\nonumber\\
&&+921 t^2 s+385 t^3\big{)} m_c^{22}+1048576 (s+t)^2
\big{(}4155 s^5+19680 t s^4+36066 t^2 s^3+34246 t^3 s^2\nonumber\\
&&+17160 t^4 s+3795 t^5\big{)} m_c^{20}-262144
\big{(}9735 s^8+69804 t s^7+219507 t^2 s^6+399776 t^3 s^5\nonumber\\
&&+466128 t^4 s^4+359956 t^5 s^3+181147 t^6 s^2+54864 t^7 s
+7755 t^8\big{)} m_c^{18}+65536 \big{(}15750 s^9\nonumber\\
&&+119943 t s^8+411033 t^2 s^7+838246 t^3 s^6+1131460 t^4 s^5
+1057300 t^5 s^4+688706 t^6 s^3\nonumber\\
&&+304133 t^7 s^2+83403 t^8 s+10890 t^9\big{)} m_c^{16}
-32768 \big{(}8955 s^{10}+72307 t s^9+267272 t^2 s^8\nonumber\\
&&+600188 t^3 s^7+913993 t^4 s^6+992452 t^5 s^5+782713 t^6 s^4
+445538 t^7 s^3+177082 t^8 s^2\nonumber\\
&&+44767 t^9 s+5445 t^{10}\big{)} m_c^{14}+4096 \big{(}14235 s^{11}
+121663 t s^{10}+482477 t^2 s^9+1181531 t^3 s^8\nonumber\\
&&+1997582 t^4 s^7+2459484 t^5 s^6+2258884 t^6 s^5+1553862 t^7 s^4
+791411 t^8 s^3+287757 t^9 s^2\nonumber\\
&&+67783 t^{10} s+7755 t^{11}\big{)} m_c^{12}-1024 \big{(}7575 s^{12}
+69080 t s^{11}+295378 t^2 s^{10}+788288 t^3 s^9\nonumber\\
&&+1469633 t^4 s^8+2024628 t^5 s^7+2117532 t^6 s^6+1697668 t^7 s^5
+1040113 t^8 s^4+480268 t^9 s^3\nonumber\\
&&+160638 t^{10} s^2+35180 t^{11} s+3795 t^{12}\big{)}
m_c^{10}+256 \big{(}2415 s^{13}+24464 t s^{12}+115885 t^2 s^{11}\nonumber\\
&&+342339 t^3 s^{10}+708615 t^4 s^9+1090700 t^5 s^8+1287226 t^6 s^7
+1180426 t^7 s^6+842220 t^8 s^5\nonumber\\
&&+464375 t^9 s^4+194239 t^{10} s^3+58985 t^{11} s^2+11684 t^{12} s
+1155 t^{13}\big{)} m_c^8-64 \big{(}345 s^{14}\nonumber\\
&&+4528 t s^{13}+25895 t^2 s^{12}+89188 t^3 s^{11}+211442 t^4 s^{10}
+369630 t^5 s^9+495006 t^6 s^8+517164 t^7 s^7\nonumber\\&&
+424126 t^8 s^6+271810 t^9 s^5+134322 t^{10} s^4+49688 t^{11} s^3
+12995 t^{12} s^2+2128 t^{13} s+165 t^{14}\big{)} m_c^6\nonumber\\
&&+16 s t (s+t) \big{(}259 s^{12}+2323 t s^{11}+9736 t^2 s^{10}
+25649 t^3 s^9+47816 t^4 s^8+66583 t^5 s^7+71232 t^6 s^6\nonumber\\
&&+58883 t^7 s^5+37456 t^8 s^4+17829 t^9 s^3+6056 t^{10} s^2+1303 t^{11} s
+139 t^{12}\big{)} m_c^4-4 s^2 t^2 (s+t)^2  \nonumber\\
&&\left(s^2+t s+t^2\right)^2 \left(87 s^6+372 t s^5+746 t^2 s^4
+854 t^3 s^3+666 t^4 s^2+292 t^5 s+67 t^6\right) m_c^2\nonumber\\
&&+13 s^3 t^3 (s+t)^3 \left(s^2+t s+t^2\right)^4\Big{]}
\end{eqnarray}

$g+g\to c\bar{c}({}^3\!P_1^{[1]})+g$:


\begin{eqnarray}
&&|\overline{N}_{gg}({}^3\!P_1^{[1]})|^2
=\frac{16 \pi ^3 \alpha_{s}^{3}}
{45 m_c^5 \left(s-4 m_c^2\right)^5
\left(4 m_c^2-t\right)^5 (s+t)^5}\nonumber\\
&&\times\Big{[}2097152 m_c^{20} (s+t)^2 \left(s^2+t^2\right)-131072 m_c^{18}
\left(11 s^5-35 s^4 t-40 s^3 t^2+5 s t^4+11 t^5\right)\nonumber\\
&&-32768 m_c^{16} \left(125 s^6+864 s^5 t+1419 s^4 t^2+1164 s^3 t^3+739 s^2
t^4+404 s t^5+85 t^6\right)\nonumber\\
&&+8192 m_c^{14} (818 s^7+4579 s^6 t+9054 s^5 t^2+9591 s^4 t^3
+7291 s^3 t^4+4914 s^2 t^5+2379 s t^6\nonumber\\
&&+538 t^7)-2048 m_c^{12} (2106 s^8+11940 s^7 t+27483 s^6 t^2+35428 s^5 t^3
+31746 s^4 t^4+23628 s^3 t^5\nonumber\\
&&+14863 s^2 t^6+6300 s t^7+1306 t^8)+512m_c^{10} (2905 s^9+17675 s^8 t
+46718 s^7 t^2+71799 s^6 t^3\nonumber\\
&&+75989 s^5 t^4+63429 s^4 t^5+44939 s^3 t^6+25058 s^2 t^7+9275 s t^8+1705 t^9)\nonumber\\
&&-128 m_c^8 (2265 s^{10}+15280 s^9 t+46309 s^8 t^2+84008 s^7 t^3+
105380 s^6 t^4+101476 s^5 t^5\nonumber\\
&&+79720 s^4t^6+51188 s^3 t^7+24729 s^2 t^8+7940 s t^9+1265 t^{10})
+32 m_c^6 (944 s^{11}+7381 s^{10} t\nonumber\\
&&+26060 s^9 t^2+56013 s^8t^3+83898 s^7 t^4+95046 s^6 t^5+85066 s^5 t^6+61038 s^4 t^7+34133 s^3 t^8\nonumber\\
&&+14020 s^2 t^9+3821 s t^{10}+504 t^{11})-8 m_c^4
   (164 s^{12}+1700 s^{11} t+7455 s^{10} t^2+19660 s^9 t^3\nonumber\\
&&+35909 s^8 t^4+48760 s^7 t^5+51084 s^6 t^6+41660 s^5 t^7+26329 s^4 t^8+12440 s^3
   t^9+4175 s^2 t^{10}\nonumber\\&&
+900 s t^{11}+84 t^{12})+2 m_c^2 s t (s+t) \left(s^2+s t+t^2\right)^2 (106 s^6+490 s^5 t+981 s^4
   t^2+1092 s^3 t^3\nonumber\\
&&+821 s^2 t^4+330 s t^5+66 t^6)-11 s^2 t^2 (s+t)^2 \left(s^2+s t+t^2\right)^4\Big{]}
\end{eqnarray}

$g+g\to c\bar{c}({}^3\!P_2^{[1]})+g$:


\begin{eqnarray}
&&|\overline{N}_{gg}({}^3\!P_2^{[1]})|^2
=\frac{16 \pi ^3 \alpha_s^3}
{225 m_c^5 s t \left(4 m_c^2-s\right)^5
\left(4 m_c^2-t\right)^5 (s+t)^5 \left(-4 m_c^2+s+t\right)}\nonumber\\
&&\times\Big{[}100663296 (s+t)^3 \left(15 s^2+38 t s+15 t^2\right)
m_c^{24}-8388608 (s+t)^2 (315 s^4+1312 ts^3\nonumber\\
&&+2010 t^2 s^2+1252 t^3 s+315 t^4) m_c^{22}+524288 (s+t)
(4620 s^6+25287 t s^5+58998 t^2 s^4\nonumber\\
&&+73262 t^3 s^3+53438 t^4 s^2+21927 t^5 s+4140 t^6)
m_c^{20}-262144 (5550 s^8+36807 t s^7\nonumber\\
&&+109272 t^2 s^6+189017 t^3 s^5+212562 t^4 s^4+162877 t^5
 s^3+84442 t^6 s^2+27387 t^7 s\nonumber\\
&&+4230 t^8)m_c^{18}+32768 (18360 s^9+126189 t s^8
+389652 t^2 s^7+714467 t^3 s^6+879782 t^4 s^5\nonumber\\
&&+782522 t^5 s^4+517507 t^6 s^3+248752 t^7 s^2+78429 t^8 s+11880 t^9)
m_c^{16}-8192 (21240 s^{10}\nonumber\\
&&+155344 t s^9+504167 t^2 s^8+967724 t^3 s^7+1252087 t^4 s^6
+1199524 t^5 s^5+911047 t^6 s^4\nonumber\\
&&+558524 t^7 s^3+262447 t^8 s^2+81544 t^9 s+11880 t^{10})
 m_c^{14}+2048 (17100 s^{11}+136829 t s^{10}\nonumber\\
&&+477883 t^2 s^9+972667 t^3 s^8+1311058 t^4 s^7+1293789 t^5 s^6
+1038389 t^6 s^5+732018 t^7 s^4\nonumber\\&&
+446587 t^8 s^3+210243 t^9 s^2+63149 t^{10} s+8460 t^{11})
m_c^{12}-512 (9180 s^{12}+82966 t s^{11}\nonumber\\
&&+321992 t^2 s^{10}+716914 t^3 s^9+1033387 t^4 s^8+1055592 t^5 s^7
+855762 t^6 s^6+634352 t^7 s^5\nonumber\\
&&+454727 t^8 s^4+283274 t^9 s^3+129252 t^{10} s^2+35566
   t^{11} s+4140 t^{12}) m_c^{10}+128 (2940 s^{13}\nonumber\\
&&+31855 t s^{12}+143846 t^2 s^{11}+365607 t^3 s^{10}+592170 t^4 s^9+663334
   t^5 s^8+565754 t^6 s^7\nonumber\\
&&+427154 t^7 s^6+326754 t^8 s^5+239410 t^9 s^4+141047 t^{10} s^3
+57346 t^{11} s^2+13375 t^{12} s \nonumber\\
&&+1260 t^{13}) m_c^8-32 (420 s^{14}+6548 t s^{13}
+37651 t^2 s^{12}+116096 t^3 s^{11}+224575 t^4 s^{10}\nonumber\\
&&+299040 t^5 s^9+299220 t^6 s^8+251484 t^7
   s^7+199880 t^8 s^6+152480 t^9 s^5+99555 t^{10} s^4\nonumber\\
&&+48976 t^{11} s^3+16051 t^{12} s^2+2828 t^{13} s+180 t^{14}) m_c^6
+8 s t (s+t) (464 s^{12}+4070 t s^{11}\nonumber\\
&&+15035 t^2 s^{10}+32905 t^3 s^9+49483 t^4 s^8+56786 t^5 s^7
+54168 t^6 s^6+44806 t^7 s^5+32183 t^8 s^4\nonumber\\
&&+18405 t^9 s^3+7675 t^{10} s^2+2030 t^{11} s+224 t^{12})
m_c^4-2 s^2 t^2 (s+t)^2 \left(s^2+t s+t^2\right)^2(138 s^6\nonumber\\
&&+552 t s^5+943 t^2 s^4+922 t^3 s^3+783 t^4 s^2+392 t^5 s+98 t^6)
m_c^2+7 s^3 t^3 (s+t)^3 \left(s^2+t s+t^2\right)^4\Big{]}\nonumber\\
\end{eqnarray}

$g+g\to c\bar{c}({}^3\!S_1^{[8]})+g$:


\begin{eqnarray}
&&|\overline{N}_{gg}({}^3\!S_1^{[8]})|^2
=\frac{\pi ^3 \alpha_{s}^{3}}{54 m_c^5
\left(s-4 m_c^2\right)^3 \left(4 m_c^2-t\right)^3 (s+t)^3}\nonumber\\
&&\times\Big{[}16384 m_c^{14} \left(87 s^2+22 s t+87 t^2\right)-4096 m_c^{12}
\left(14 s^3-449 s^2 t-221 s t^2+14 t^3\right)\nonumber\\
&&-2048 m_c^{10} \left(480 s^4+1969 s^3 t+2182 s^2 t^2+1384 s t^3
+399 t^4\right)+256 m_c^8 (2910 s^5 \nonumber\\
&&+11616 s^4   t+17509 s^3 t^2+15433 s^2 t^3+8178 s t^4+2100 t^5)-64 m_c^6
(4048 s^6+17148 s^5 t \nonumber\\
&&+31739 s^4 t^2+35722 s^3 t^3+25841 s^2t^4+11412 s t^5+2590 t^6)+16 m_c^4
(2754 s^7+12968 s^6 t\nonumber\\
&&+28779 s^5 t^2+39352 s^4 t^3+35950 s^3 t^4+22137 s^2 t^5+8270 st^6+1620 t^7)\nonumber\\
&&-108 m_c^2 \left(s^2+s t+t^2\right) \left(27 s^6+131 s^5 t+269 s^4 t^2
+302 s^3 t^3+227 s^2 t^4+89 s t^5+15t^6\right)\nonumber\\
&&+297 s t (s+t) \left(s^2+s t+t^2\right)^3\Big{]}
\end{eqnarray}

$g+g\to c\bar{c}({}^1\!S_0^{[8]})+g$:

\begin{eqnarray}
&&|\overline{N}_{gg}({}^1\!S_0^{[8]})|^2=\frac{5 \pi ^3 \alpha_s^3}
{3 m_c^3 s t \left(4 m_c^2-s\right)^3 \left(4 m_c^2
-t\right)^3 (s+t)^3 \left(-4 m_c^2+s+t\right)}\nonumber\\
&&\times\Big{[}65536 m_c^{16} \left(5 s^3+20 s^2 t+8 s t^2+5 t^3\right)
-16384 m_c^{14} (25 s^4+116 s^3 t+110 s^2 t^2+56 s t^3\nonumber\\
&&+25 t^4 )+12288 m_c^{12} \left(28 s^5+129 s^4 t+189 s^3 t^2
+137 s^2 t^3+65 s t^4+20 t^5\right)\nonumber\\
&&-2048 m_c^{10} \left(87 s^6+426 s^5 t+801 s^4 t^2+806 s^3 t^3
+501 s^2 t^4+204 s t^5+45 t^6\right)\nonumber\\
&&+256 m_c^8 \left(222 s^7+1182 s^6 t+2653 s^5 t^2+3385 s^4 t^3
+2749 s^3 t^4+1501 s^2 t^5+522 s t^6+90 t^7\right)\nonumber\\
&&-64 m_c^6 (180 s^8+1020 s^7 t+2591 s^6 t^2+3964 s^5t^3
+4006 s^4 t^4+2812 s^3 t^5+1355 s^2 t^6\nonumber\\
&&+408 s t^7+60 t^8)+16 m_c^4 (85 s^9+516 s^8 t+1456 s^7 t^2
+2561 s^6 t^3+3096 s^5 t^4+2688 s^4 t^5\nonumber\\
&&+1673 s^3 t^6+724 s^2 t^7+192 s t^8+25 t^9)-4 m_c^2
\left(s^2+s t+t^2\right)^2 (17 s^6+88 s^5 t+167 s^4 t^2\nonumber\\
&&+168 s^3 t^3+119 s^2 t^4+40 s t^5+5 t^6)+5 s t (s+t) \left(s^2+s t+t^2\right)^4\Big{]}
\end{eqnarray}

$g+g\to c\bar{c}({}^3\!P_J^{[8]})+g$:

\begin{eqnarray}
&&|\overline{N}_{gg}({}^3\!P_J^{[8]})|^2= \frac{\pi ^3 \alpha_s^3}
{m_c^5 s t \left(s-4 m_c^2\right)^4 \left(t
-4 m_c^2\right)^4 (s+t)^4 \left(-4 m_c^2+s+t\right)}\nonumber\\
&&\times\Big{[}1048576 m_c^{20} \left(115 s^4+435 s^3 t+564 s^2 t^2
+295 s t^3+115 t^4\right)-524288 m_c^{18}(345 s^5 \nonumber\\
&&+1571 s^4 t+2683 s^3 t^2+2148 s^2 t^3+1156 s t^4+345 t^5)
+65536 m_c^{16} (2235 s^6+11102 s^5 t\nonumber\\
&&+22740 s^4 t^2+23720 s^3 t^3+16400 s^2 t^4+7932 s t^5
+1955 t^6)-16384 m_c^{14} (4710 s^7\nonumber\\
&&+24932 s^6 t+57177 s^5 t^2+71731 s^4 t^3+60491 s^3 t^4
+37287 s^2 t^5+16102 s t^6+3450 t^7 )\nonumber\\
&&+4096 m_c^{12} (6660 s^8+38018 s^7 t+94927 s^6 t^2
+134978 s^5 t^3+132094 s^4 t^4+97708 s^3 t^5\nonumber\\
&&+54647 s^2 t^6+21708 s t^7+4140 t^8)-1024 m_c^{10}
(6390 s^9+39787 s^8 t+107560 s^7 t^2\nonumber\\
&&+168509 s^6 t^3+183028 s^5 t^4+154598 s^4 t^5+104319 s^3 t^6
+54540 s^2 t^7+20307 s t^8+3450 t^9)\nonumber\\
&&+256 m_c^8 (4055 s^{10}+27855 s^9 t+82334 s^8 t^2
+142320 s^7 t^3+169197 s^6 t^4+156936 s^5 t^5\nonumber\\
&&+120977 s^4 t^6+77050 s^3 t^7+38034 s^2 t^8+13095 s t^9
+1955 t^{10})-64 m_c^6 (1530 s^{11}\nonumber\\
&&+12009 s^{10} t+40145 s^9 t^2+79031 s^8 t^3+106482 s^7 t^4
+110003 s^6 t^5+93903 s^5 t^6+68322 s^4 t^7\nonumber\\
&&+40531 s^3 t^8+18105 s^2 t^9+5449 s t^{10}+690 t^{11})
+16 m_c^4 (255 s^{12}+2610 s^{11} t+10798 s^{10} t^2\nonumber\\
&&+26205 s^9 t^3+43547 s^8 t^4+54320 s^7 t^5+53906 s^6 t^6
+43520 s^5 t^7+28607 s^4 t^8+14485 s^3 t^9\nonumber\\
&&+5278 s^2 t^{10}+1230 s t^{11}+115 t^{12})-4 m_c^2 s t
(s+t) \left(s^2+s t+t^2\right)^2 (150 s^6+696 s^5 t\nonumber\\
&&+1299 s^4 t^2+1358 s^3 t^3+1059 s^2 t^4+456 s t^5+90 t^6)
+31 s^2 t^2 (s+t)^2 \left(s^2+s t+t^2\right)^4\Big{]}
\end{eqnarray}

$g+g\to c\bar{c}({}^3\!S_1^{[8]},{}^3\!D_1^{[8]})+g$:

\begin{eqnarray}
&&|\overline{N}_{gg}({}^3\!S_1^{[8]},{}^3\!D_1^{[8]})|^2
=-\frac{\pi ^3 \alpha_{s}^{3}} {108 \sqrt{15} m_c^5
\left(s-4 m_c^2\right)^4 \left(t-4 m_c^2\right)^4 (s+t)^4} \nonumber\\
&&\times\Big{[}-524288 m_c^{20} \left(4463 s^2
+25840 s t+5057 t^2\right)+65536 m_c^{18} (s+t) (7767s^2+145760 s t\nonumber\\
&&+13689 t^2)+16384 m_c^{16} \left(40922 s^4+83203 s^3 t
-78401 s^2 t^2+46753 s t^3+25451 t^4\right)\nonumber\\
&&-4096 m_c^{14} (s+t) \left(41543 s^4+579922 s^3 t
+620206 s^2 t^2+525220 s t^3+20753 t^4\right)-3072 m_c^{12}\nonumber\\
&&\left(40919 s^6-153453 s^5 t-767319 s^4 t^2-1075115 s^3 t^3-737820 s^2 t^4
-133812 s t^5+45008 t^6\right)\nonumber\\
&&+256 m_c^{10} (s+t) (302824 s^6+248112 s^5 t-1111219 s^4 t^2
-1766115 s^3 t^3-1102093 s^2 t^4\nonumber\\
&&+250353 s t^5+304390 t^6 )-64 m_c^8 (287151 s^8
+1147982 s^7 t+1476785 s^6 t^2+462324 s^5 t^3\nonumber\\
&&-288201 s^4 t^4+391881 s^3 t^5+1418006 s^2 t^6+1132457 s t^7
+286503 t^8)+16 m_c^6 (s+t)\nonumber\\
&&(136080 s^8+645678 s^7 t+1174264 s^6 t^2+1234125 s^5 t^3
+1166106 s^4 t^4+1190394 s^3 t^5\nonumber\\
&&+1143268 s^2 t^6+640305 s t^7+136080 t^8)-4 m_c^4
(27216 s^{10}+226800 s^9 t+774349 s^8 t^2\nonumber\\
&&+1531962 s^7 t^3+2158036 s^6 t^4+2377567 s^5 t^5
+2135518 s^4 t^6+1512603 s^3 t^7+768949 s^2 t^8\nonumber\\
&&+226800 s t^9+27216 t^{10})+m_c^2 s t (s+t)
(18144 s^8+104976 s^7 t+274390 s^6 t^2+449444 s^5 t^3\nonumber\\
&&+528929 s^4 t^4+449390 s^3 t^5+274363 s^2 t^6+104976 s t^7+18144 t^8)\nonumber\\
&&-1620 s^2 t^2 (s+t)^2 \left(s^2+s t+t^2\right)^3\Big{]}
\end{eqnarray}


$q(\bar{q})+g\to c\bar{c}({}^3\!P_0^{[1]})+q(\bar{q})$:


\begin{eqnarray}
&&|\overline{N}_{q(\bar{q})g}({}^3\!P_0^{[1]})|^2
=\frac{16 \pi ^3 \alpha_{s}^{3}}
{405 m_c^5 t \left(4 m_c^2-s\right) \left(4 m_c^2- t\right)^5}
\left(12 m_c^2-t\right) \Big{[}87040 m_c^{10}-256 m_c^8 \nonumber\\
&& (285 s+278 t)+256 m_c^6 \left(115 s^2 +166 s t
+81 t^2\right)-32 m_c^4 \big{(}115 s^3+311 s^2 t + 243 s t^2+77 t^3\big{)}\nonumber\\
&& +4 m_c^2 t \left(96 s^3+188 s^2 t+120 s t^2+23 t^3\right)
-13 s t^2 \left(2 s^2+2 s t+t^2\right)\Big{]}
\end{eqnarray}

$q(\bar{q})+g\to c\bar{c}({}^3\!P_1^{[1]})+q(\bar{q})$:


\begin{eqnarray}
&&|\overline{N}_{q(\bar{q})g}({}^3\!P_1^{[1]})|^2
=\frac{32 \pi ^3 \alpha_s^3}
{405 m_c^5 \left(4 m_c^2-s\right)
\left(4 m_c^2-t\right)^5} \Big{[}256 m_c^8 (124 s+21 t)
-64 m_c^6  \nonumber\\&&(288 s^2
+341 s t+63 t^2)+16 m_c^4
(164 s^3+456 s^2 t+321 s t^2+63 t^3)-4 m_c^2 t(106 s^3\nonumber\\
&&+190 s^2 t+115 s t^2+21 t^3)+11 s t^2 \left(2 s^2+2 s t+t^2\right)\Big{]}
\end{eqnarray}

$q(\bar{q})+g\to c\bar{c}({}^3\!P_2^{[1]})+q(\bar{q})$:


\begin{eqnarray}
&&|\overline{N}_{q(\bar{q})g}({}^3\!P_2^{[1]})|^2
=\frac{32 \pi ^3 \alpha_s^3}
{2025 m_c^5 t \left(4 m_c^2-s\right) \left(4 m_c^2-t\right)^5}
 \Big{[}614400 m_c^{12}-34816 m_c^{10}(15 s\nonumber\\
&&+13 t)+256 m_c^8 \left(840 s^2+1386 s t+443 t^2\right)
-64 m_c^6 \left(420 s^3+1708 s^2 t+1315 s t^2+177 t^3\right)\nonumber\\
&&+16 m_c^4 t \left(464 s^3+1060 s^2 t+519 s t^2+43
   t^3\right)-4 m_c^2 t^2 \left(138 s^3+206 s^2 t+87 s t^2+17 t^3\right)\nonumber\\
&&+7 s t^3 \left(2 s^2+2 s t+t^2\right)\Big{]}
\end{eqnarray}

$q(\bar{q})+g\to c\bar{c}({}^3\!S_1^{[8]})+q(\bar{q})$:

\begin{eqnarray}
&&|\overline{N}_{q(\bar{q})g}({}^3\!S_1^{[8]})|^2=\frac{\pi ^3 \alpha_s^3}
{81 m_c^5 s t \left(4 m_c^2-s\right) (s+t)^3}
\Big{[}128 m_c^6 \left(20 s^3+69 s^2 t-39 s t^2+20 t^3\right)\nonumber\\
&&-32 m_c^4 \left(40 s^4+113 s^3 t+27 s^2 t^2+10 s t^3+20 t^4\right)
+4 m_c^2 (108 s^5+193 s^4 t+41 s^3 t^2\nonumber\\
&&+225 s^2 t^3+s t^4+20 t^5)-11 s \left(4 s^5+3 s^4 t+7 s^3
   t^2+7 s^2 t^3+3 s t^4+4 t^5\right)\Big{]}
\end{eqnarray}

$q(\bar{q})+g\to c\bar{c}({}^1\!S_0^{[8]})+q(\bar{q})$:

\begin{eqnarray}
&&|\overline{N}_{q(\bar{q})g}({}^1\!S_0^{[8]})|^2=\frac{10 \pi ^3 \alpha_s^3
\left[m_c^2 \left(44 s^3+92 s^2 t-4 s t^2+44 t^3\right)
-5 s \left(s^3+s^2 t+s t^2+t^3\right)\right]}{27 m_c^3
\left(s-4 m_c^2\right) (s+t)^3 \left(-4 m_c^2+s+t\right)}
\end{eqnarray}

$q(\bar{q})+g\to c\bar{c}({}^3\!P_J^{[8]})+q(\bar{q})$:

\begin{eqnarray}
&&|\overline{N}_{q(\bar{q})g}({}^3\!P_J^{[8]})|^2=\frac{2 \pi ^3 \alpha_s^3}
{9 m_c^5 \left(s-4 m_c^2\right) (s+t)^4 \left(-4 m_c^2+s+t\right)}
\Big{[}512 m_c^6 \left(5 s^2+26 s t+25 t^2\right)\nonumber\\
&&+64 m_c^4 \left(s^3-23 s^2 t-111 s t^2-19 t^3\right)+4 m_c^2 (s+t)
\left(57 s^3+169 s^2 t-3 s t^2+61 t^3\right)\nonumber\\
&&-31 s (s+t)^2 \left(s^2+t^2\right)\Big{]}
\end{eqnarray}

$q(\bar{q})+g\to c\bar{c}({}^3\!S_1^{[8]},{}^3\!D_1^{[8]})+q(\bar{q})$:

\begin{eqnarray}
&&|\overline{N}_{q(\bar{q})g}({}^3\!S_1^{[8]},{}^3\!D_1^{[8]})|^2
=-\frac{2 \pi ^3 \alpha_{s}^{3}}
{27 \sqrt{15} m_c^5 s t (s+t)^4}
\Big{[}128 m_c^4 \left(5 s^4+2 s^3 t+21 s^2 t^2+2 s t^3+5 t^4\right)\nonumber\\
&&-32 m_c^2 \left(5 s^5+16 s^4 t+14 s^3 t^2+14 s^2 t^3+16 s t^4+5 t^5\right)\nonumber\\
&&+5 (s+t)^2 \left(4 s^4-s^3 t+8 s^2 t^2-s t^3+4 t^4\right)\Big{]}
\end{eqnarray}


$\bar{q}+q\to c\bar{c}({}^3\!P_0^{[1]})+g$:


\begin{eqnarray}
&&|\overline{N}_{\bar{q}q}({}^3\!P_0^{[1]})|^2
=-\frac{128 \pi ^3 \alpha_{s}^{3}\left(12 m_c^2-s\right)
\left(880 m_c^4-112 m_c^2 s+13 s^2\right) } {3645 m_c^5 s
\left(4 m_c^2-s\right)^5}
\nonumber\\ &&\times\big{[}2 t \left(s-4 m_c^2\right)+\left(s-4 m_c^2\right)^2+2 t^2\big{]}
\end{eqnarray}

$\bar{q}+q\to c\bar{c}({}^3\!P_1^{[1]})+g$:


\begin{eqnarray}
&&|\overline{N}_{\bar{q}q}({}^3\!P_1^{[1]})|^2
=\frac{256 \pi ^3 \alpha_s^3}
{1215 m_c^5 \left(s-4 m_c^2\right)^5}
 \Big{[}64 m_c^6 (31 s+84 t)-16 m_c^4 (73 s^2+150 s t\nonumber\\
&&+84 t^2)+4 m_c^2 s \left(53 s^2+88 s t+66 t^2\right)
-11 s^2 \left(s^2+2 s t+2 t^2\right)\Big{]}
\end{eqnarray}

$\bar{q}+q\to c\bar{c}({}^3\!P_2^{[1]})+g$:


\begin{eqnarray}
&&|\overline{N}_{\bar{q}q}({}^3\!P_2^{[1]})|^2
=\frac{256 \pi ^3 \alpha_s^3}{6075 m_c^5 s \left(s-4 m_c^2\right)^5}
 \Big{[}92160 m_c^{10}-512 m_c^8 (71 s+90 t)+64 m_c^6 \nonumber\\
&&(33 s^2+404 s t+180 t^2)-112 m_c^4 s \left(s^2+46 s t+32 t^2\right)
+4 m_c^2 s^2 \left(33 s^2+112 s t+98 t^2\right)\nonumber\\
&&-7 s^3 \left(s^2+2 s t+2 t^2\right)\Big{]}
\end{eqnarray}

$\bar{q}+q\to c\bar{c}({}^3\!S_1^{[8]})+g$:

\begin{eqnarray}
&&|\overline{N}_{\bar{q}q}({}^3\!S_1^{[8]})|^2=\frac{8 \pi ^3 \alpha_s^3}
{243 m_c^5 t \left(4 m_c^2-s\right)^3 \left(-4 m_c^2+s+t\right)}
\Big{[}45056 m_c^{10}\nonumber\\
&&-256 m_c^8 (84 s+205 t)+64 m_c^6 \left(128 s^2+496 s t
+475 t^2\right)-16 m_c^4 (224 s^3+780 s^2 t \nonumber\\
&&+1029 s t^2+540 t^3)+4 m_c^2 \left(180 s^4+676 s^3 t+1155 s^2 t^2
+936 s t^3+270 t^4\right)\nonumber\\
&&-11 s \left(4 s^4+17 s^3 t+35 s^2 t^2+36 s t^3+18 t^4\right)\Big{]}
\end{eqnarray}

$\bar{q}+q\to c\bar{c}({}^1\!S_0^{[8]})+g$:

\begin{eqnarray}
&&|\overline{N}_{\bar{q}q}({}^1\!S_0^{[8]})|^2=-\frac{400 \pi ^3 \alpha_s^3
\left[16 m_c^4-8 m_c^2 (s+t)+s^2+2 s t+2 t^2\right]}{81 m_c^3 s
\left( s-4 m_c^2\right)^2}
\end{eqnarray}

$\bar{q}+q\to c\bar{c}({}^3\!P_J^{[8]})+g$:

\begin{eqnarray}
&&|\overline{N}_{\bar{q}q}({}^3\!P_J^{[8]})|^2=\frac{-16 \pi ^3 \alpha_s^3}
{27 m_c^5 s \left(s-4 m_c^2\right)^4}\Big{[}29440 m_c^8
-128 m_c^6 (68 s+115 t)+32 m_c^4 (59 s^2\nonumber\\
&&+205 s t+115 t^2)-8 m_c^2 s \left(64 s^2+121 s t+90 t^2\right)
+31 s^2 \left(s^2+2 s t+2 t^2\right)\Big{]}
\end{eqnarray}

$\bar{q}+q\to c\bar{c}({}^3\!S_1^{[8]},{}^3\!D_1^{[8]})+g$:

\begin{eqnarray}
&&|\overline{N}_{\bar{q}q}({}^3\!S_1^{[8]},{}^3\!D_1^{[8]})|^2
=-\frac{16 \pi ^3 \alpha_{s}^{3}}
{81 \sqrt{15} m_c^5 t \left(s-4 m_c^2\right)^4
\left(-4 m_c^2+s+t\right)} \Big{[}81920 m_c^{12}\nonumber\\
&&-1024 m_c^{10} (80 s+157 t)+256 m_c^8
\left(140 s^2+569 s t+535 t^2\right)-128 m_c^6 (80 s^3+425 s^2 t\nonumber\\
&&+710 s t^2+378 t^3 )+32 m_c^4 \left(70 s^4+353 s^3 t
+705 s^2 t^2+630 s t^3+189 t^4\right)-4 m_c^2 s (80 s^4 \nonumber\\
&&+353 s^3 t+700 s^2 t^2+684 s t^3+252 t^4)+5 s^2
\left(4 s^4+17 s^3 t+35 s^2 t^2+36 s t^3+18
   t^4\right)\Big{]}
\end{eqnarray}


\begin{thebibliography}{99}

\bibitem{Caswell:1985ui}
  W.~E.~Caswell and G.~P.~Lepage,
  Phys.\ Lett.\  B {\bf 167}, 437 (1986).

\bibitem{Bodwin:1994jh}
  G.~T.~Bodwin, E.~Braaten, and G.~P.~Lepage,
  Phys.\ Rev.\  D {\bf 51}, 1125 (1995);
  {\bf 55}, 5853(E) (1997)
  [hep-ph/9407339].

\bibitem{Lepage:1992tx}
  G.~P.~Lepage, L.~Magnea, C.~Nakhleh, U.~Magnea, and K.~Hornbostel,
  Phys.\ Rev.\ D {\bf 46}, 4052 (1992)
  [hep-lat/9205007].

\bibitem{Campbell:2007ws}
  J.~Campbell, F.~Maltoni, and F.~Tramontano,
  Phys.\ Rev.\ Lett.\  {\bf 98}, 252002 (2007)
  [hep-ph/0703113].

\bibitem{Gong:2008sn}
  B.~Gong and J.-X.~Wang,
  Phys.\ Rev.\ Lett.\ {\bf 100}, 232001 (2008)
  [arXiv:0802.3727 [hep-ph]].

\bibitem{Butenschoen:2010rq}
  M.~Butensch\"on and B.~A.~Kniehl,
  Phys.\ Rev.\ Lett.\  {\bf 106}, 022003 (2011)
  [arXiv:1009.5662 [hep-ph]].

\bibitem{Ma:2010yw}
  Y.-Q.~Ma, K.~Wang, and K.-T.~Chao,
  Phys.\ Rev.\ Lett.\  {\bf 106}, 042002 (2011)
  [arXiv:1009.3655 [hep-ph]].

\bibitem{Ma:2010vd}
  Y.-Q.~Ma, K.~Wang, and K.-T.~Chao,
  Phys.\ Rev.\ D {\bf 83}, 111503(R) (2011)
  [arXiv:1002.3987 [hep-ph]].

\bibitem{Butenschoen:2013pxa}
  M.~Butenschoen, Z.-G.~He, and B.~A.~Kniehl,
  Phys.\ Rev.\ D {\bf 88}, 011501(R) (2013)
  [arXiv:1303.6524 [hep-ph]].

\bibitem{Butenschoen:2011yh}
  M.~Butenschoen and B.~A.~Kniehl,
  Phys.\ Rev.\ D {\bf 84}, 051501(R) (2011)
  [arXiv:1105.0820 [hep-ph]].

\bibitem{Butenschoen:2012px}
  M.~Butenschoen and B.~A.~Kniehl,
  Phys.\ Rev.\ Lett.\  {\bf 108}, 172002 (2012)
  [arXiv:1201.1872 [hep-ph]].

\bibitem{Chao:2012iv}
  K.-T.~Chao, Y.-Q.~Ma, H.-S.~Shao, K.~Wang, and Y.-J.~Zhang,
  Phys.\ Rev.\ Lett.\ {\bf 108}, 242004 (2012)
  [arXiv:1201.2675 [hep-ph]].

\bibitem{Gong:2012ug}
  B.~Gong, L.-P.~Wan, J.-X.~Wang, and H.-F.~Zhang,
  Phys.\  Rev.\  Lett.\ {\bf 110}, 042002 (2013)
  [arXiv:1205.6682 [hep-ph]].

\bibitem{Chatrchyan:2013cla}
  S.~Chatrchyan {\it et al.}\  (CMS Collaboration),
  Phys.\ Lett.\ B {\bf 727}, 381 (2013)
  [arXiv:1307.6070 [hep-ex]].

\bibitem{Aaij:2013nlm}
  R.~Aaij {\it et al.}\  (LHCb Collaboration),
  Eur.\ Phys.\ J.\ C {\bf 73}, 2631 (2013)
  [arXiv:1307.6379 [hep-ex]].

\bibitem{He:2007te}
  Z.-G.~He, Y.~Fan, and K.-T.~Chao,
  Phys.\ Rev.\ D {\bf 75}, 074011 (2007)
  [hep-ph/0702239].

\bibitem{He:2009uf}
  Z.-G.~He, Y.~Fan, and K.~-T.~Chao,
  Phys.\ Rev.\ D {\bf 81}, 054036 (2010)
  [arXiv:0910.3636 [hep-ph]].

\bibitem{Jia:2009np}
  Y.~Jia,
  Phys.\ Rev.\ D {\bf 82}, 034017 (2010)
  [arXiv:0912.5498 [hep-ph]].

\bibitem{Xu:2012am}
  G.-Z.~Xu, Y.-J.~Li, K.-Y.~Liu, and Y.-J.~Zhang,
  Phys.\ Rev.\ D {\bf 86}, 094017 (2012)
  [arXiv:1203.0207 [hep-ph]].

\bibitem{Fan:2009zq}
  Y.~Fan, Y.-Q.~Ma, and K.~-T.~Chao,
  Phys.\ Rev.\ D {\bf 79}, 114009 (2009)
  [arXiv:0904.4025 [hep-ph]].

\bibitem{Martynenko:2012tf}
  A.~P.~Martynenko and A.~M.~Trunin,
  Phys.\ Rev.\ D {\bf 86}, 094003 (2012)
  [arXiv:1207.3245 [hep-ph]];
  Phys.\ Lett.\ B {\bf 723}, 132 (2013)
  [arXiv:1302.6726 [hep-ph]].

\bibitem{Li:2013csa}
  Y.-J.~Li, G.-Z.~Xu, K.-Y.~Liu, and Y.-J.~Zhang,
  J. High Energy Phys.\ {\bf 07} (2013) 051
  [arXiv:1303.1383 [hep-ph]].

\bibitem{Brambilla:2008zg}
  N.~Brambilla, A.~Vairo, and E.~Mereghetti,
  Phys.\ Rev.\ D {\bf 79}, 074002 (2009);
  {\bf 83}, 079904(E) (2011)
  [arXiv:0810.2259 [hep-ph]].

\bibitem{Kuhn:1979bb}
  J.~H.~K\"uhn, J.~Kaplan, and E.~G.~O.~Safiani,
  Nucl.\ Phys.\  {\bf B157}, 125 (1979);
  B.~Guberina, J.~H.~K\"uhn, R.~D.~Peccei, and R.~Ruckl,
  Nucl.\ Phys.\  {\bf B174}, 317 (1980);
  E.~L.~Berger and D.~Jones,
  Phys.\ Rev.\  D {\bf 23}, 1521 (1981).

\bibitem{Bodwin:2002hg}
  G.~T.~Bodwin and A.~Petrelli,
  Phys.\ Rev.\ D {\bf 66}, 094011 (2002)
  [hep-ph/0205210].

\bibitem{Kublbeck:1990xc}
  J.~K\"ublbeck, M.~B\"ohm, and A.~Denner,
  Comput.\ Phys.\ Commun.\  {\bf 60}, 165 (1990).

\bibitem{Mertig:1990an}
  R.~Mertig, M.~B\"ohm, and A.~Denner,
  Comput.\ Phys.\ Commun.\  {\bf 64}, 345 (1991).

\bibitem{Cho:1995vh}
  P.~Cho and A.~K.~Leibovich,
  Phys.\ Rev.\ D {\bf 53}, 150 (1996)
  [hep-ph/9505329];
{\bf 53}, 6203 (1996)
  [hep-ph/9511315].

\bibitem{Ko:1996xw}
  P.~Ko, J.~Lee, and H.~S.~Song,
  Phys.\ Rev.\ D {\bf 54}, 4312 (1996);
  {\bf 60}, 119902(E) (1999)
  [hep-ph/9602223].

\bibitem{Jung:1993cd}
  H.~Jung, D.~Kr\"ucker, C.~Greub, and D.~Wyler,
  Z.\ Phys.\ C {\bf 60}, 721 (1993).

\bibitem{Pumplin:2002vw}
  J.~Pumplin, D.~R.~Stump, J.~Huston, H.-L.~Lai, P.~M.~Nadolsky,
 and W.-K.~Tung,
  J. High Energy Phys.\ {\bf 07} (2002) 012
  [hep-ph/0201195].

\bibitem{Kniehl:1996we}
  B.~A.~Kniehl, G.~Kramer, and M.~Spira,
  Z.\ Phys.\ C {\bf 76}, 689 (1997)
  [hep-ph/9610267].

\bibitem{Aaron:2010gz}
  F.~D.~Aaron {\it et al.}\  (H1 Collaboration),
  Eur.\ Phys.\ J.\ C {\bf 68}, 401 (2010)
  [arXiv:1002.0234 [hep-ex]].

\bibitem{Bodwin:2007fz}
  G.~T.~Bodwin, H.~S.~Chung, D.~Kang, J.~Lee and C.~Yu,
  Phys.\ Rev.\ D {\bf 77}, 094017 (2008)
  [arXiv:0710.0994 [hep-ph]].

\bibitem{Guo:2011tz}
  H.~-K.~Guo, Y.~-Q.~Ma and K.~-T.~Chao,
  Phys.\ Rev.\ D {\bf 83}, 114038 (2011)
  [arXiv:1104.3138 [hep-ph]].

\bibitem{Beringer:1900zz}
  J.~Beringer {\it et al.}\  (Particle Data Group),
  Phys.\ Rev.\ D {\bf 86}, 010001 (2012).

\bibitem{Abe:1997jz}
  F.~Abe {\it et al.}\  (CDF Collaboration),
  Phys.\ Rev.\ Lett.\  {\bf 79}, 572 (1997);
  Phys.\ Rev.\ Lett.\  {\bf 79}, 578 (1997).

\bibitem{Aaij:2011jh}
  R.~Aaij {\it et al.}\  (LHCb Collaboration),
  Eur.\ Phys.\ J.\ C {\bf 71}, 1645 (2011)
  [arXiv:1103.0423 [hep-ex]].

\end{thebibliography}
\end{document}